\documentclass{aa}  

\usepackage{graphicx}
\usepackage{subcaption}
\usepackage{txfonts}
\usepackage{siunitx}
\usepackage{comment}
\usepackage{adjustbox,textcomp}

\begin{document}

   \title{The ALMA-ALPINE [CII] survey: The star formation history and the dust emission of star-forming galaxies at 4.5 $<$ z $<$ 6.2}

   \subtitle{}

   \titlerunning{IR characteristic emission and dust properties of star-forming galaxies at 4.5 $<$ z $<$ 6.2}

   \author{D. Burgarella
          \inst{1}
           \and
          J. Bogdanoska\inst{1}
          \and
          A. Nanni\inst{2}
          \and
          S. Bardelli\inst{3}
          \and
          M. B\'ethermin\inst{1}
          \and
          M. Boquien\inst{4}
          \and
          V. Buat\inst{1, 32}
          \and
          A. L. Faisst\inst{5}
          \and
          M. Dessauges-Zavadsky\inst{6}
          \and
          Y. Fudamoto\inst{7,8}
          \and
          S. Fujimoto\inst{9,10}
          \and
          M. Giavalisco\inst{11}
          \and
          M. Ginolfi\inst{12}
          \and
          C. Gruppioni\inst{13}
          \and
          N. P. Hathi\inst{14}
          \and
          E. Ibar\inst{15}
          \and
          G. C. Jones\inst{16,17}
          \and
          A. M. Koekemoer\inst{14}
          \and
          K. Kohno\inst{18,19}
          \and
          B. C. Lemaux\inst{20, 21}
          \and
          D. Narayanan\inst{22}
          \and
          P. Oesch\inst{6, 9, 10}
          \and
          M. Ouchi\inst{18,23,24}
          \and
          D. A. Riechers\inst{25}
          \and
          F. Pozzi\inst{3, 26}
          \and
          M. Romano\inst{2, 27, 28}
          \and
          D. Schaerer\inst{6,29}
          \and
          M. Talia\inst{3,30}
          \and
          P. Theul\'e\inst{1}
          \and
          D. Vergani\inst{3}
          \and
          G. Zamorani\inst{3}
          \and
          E. Zucca\inst{3}
          \and
          P. Cassata\inst{31}
          \and
         the ALPINE team
          }
   \institute{Aix Marseille Univ, CNRS, CNES, LAM, Marseille, France\\
              \email{denis.burgarella@lam.fr}
         \and
             National Centre for Nuclear Research, ul.Pasteura 7, 02-093 Warszawa, Poland
         \and
            INAF - Osservatorio di Astrofisica e Scienza dello Spazio di Bologna, via Gobetti 93/3 - 40129 Bologna - Italy
         \and
             Centro de Astronomía (CITEVA), Universidad de Antofagasta, Avenida Angamos 601, Antofagasta, Chile
         \and
             IPAC, California Institute of Technology, 1200 East California Boulevard, Pasadena, CA 91125, USA
          \and
             Department of Astronomy, University of Geneva, ch. des Maillettes 51, CH-1290 Versoix, Switzerland
            \and
             National Astronomical Observatory of Japan, 2-21-1, Osawa, Mitaka, Tokyo, Japan
            \and
             Research Institute for Science and Engineering, Waseda University, 3-4-1 Okubo, Shinjuku, Tokyo 169-8555, Japan
            \and
             Cosmic Dawn Center (DAWN), Jagtvej 128, DK2200, Copenhagen N, Denmark
            \and
             Niels Bohr Institute, University of Copenhagen, Lyngbyvej 2, DK2100 Copenhagen, Denmark
             \and
             University of Massachusetts, 710 N. Pleasant St, LGRB-520, Amherst, MA 01003, USA
             \and
             European Southern Observatory, Karl-Schwarzschild-Str. 2, D-85748 Garching bei München, Germany
            \and
             Istituto Nazionale di Astrofisica: Osservatorio di Astrofisica e Scienza dello Spazio di Bologna, Via Gobetti 93/3, 40129 Bologna, Italy
            \and
             Space Telescope Science Institute, 3700 San Martin Drive, Baltimore, MD 21218, USA
            \and
             Instituto de Física y Astronomía, Universidad de Valparaíso, Avda. Gran Bretaña 1111, Valparaíso, Chile
         \and
             Cavendish Laboratory, University of Cambridge, 19 J. J. Thomson Ave., Cambridge CB3 0HE, UK
         \and
             Kavli Institute for Cosmology, University of Cambridge, Madingley Road, Cambridge CB3 0HA, UK
         \and
             Institute of Astronomy, School of Science, The University of Tokyo, 2-21-1 Osawa, Mitaka, Tokyo 181-0015, Japan
         \and
             Research Center for the Early Universe, School of Science, The University of Tokyo, 7-3-1 Hongo, Bunkyo, Tokyo 113-0033, Japan
            \and
             Department of Physics, University of California, Davis, One Shields Ave., Davis, CA 95616, USA
            \and
             Gemini Observatory, NSF’s NOIRLab, 670 N. A’ohoku Place, Hilo, Hawai’i, 96720, USA
            \and
             Department of Astronomy, University of Florida, 211 Bryant Space Sciences Center, Gainesville, FL 32611 USA
             \and
             Institute for Cosmic Ray Research, The University of Tokyo, 5-1-5 Kashiwanoha, Kashiwa, Chiba 277-8582, Japan
            \and
             Kavli Institute for the Physics and Mathematics of the Universe (WPI), The University of Tokyo, 5-1-5 Kashiwanoha, Kashiwa, Chiba 277-8583, Japan
            \and
             Cornell University, Space Sciences Building, Ithaca, NY 14853, USA
            \and
             Dipartimento di Fisica e Astronomia, Università degli Studi di Bologna, Via P. Gobetti 93/2, I-40129 Bologna, Italy
            \and
             Dipartimento di Fisica e Astronomia, Universit\`a di Padova, Vicolo dell'Osservatorio 3, I-35122, Padova, Italy
            \and
            INAF - Osservatorio Astronomico di Padova, Vicolo dell'Osservatorio 5, I-35122, Padova, Italy
            \and
             Institut de Recherche en Astrophysique et Planétologie - IRAP, CNRS, Université de Toulouse, UPS-OMP, 14, avenue E. Belin, F31400 Toulouse, France
            \and
             University of Bologna - Department of Physics and Astronomy “Augusto Righi” (DIFA), Via Gobetti 93/2, I-40129, Bologna, Italy
            \and
             Dipartimento di Fisica e Astronomia Galileo Galilei Università degli Studi di Padova, Vicolo dell’Osservatorio 3, 35122 Padova Italy
            \and
             Institut Universitaire de France, IUF, France
        }
   \date{Received November XX, 2021; accepted ...}
   
 
  \abstract {
   {}
  {Star-forming galaxies are composed of various types of galaxies. However, the luminosity functions at z $\gtrsim$ 4 - 5 suggest that most galaxies have a relatively low stellar mass ($\log M_{star} \sim 10$) and a low dust attenuation (A$_{FUV} \sim 1.0$). The physical properties of these objects are quite homogeneous. We used an approach where we combined their rest-frame far-infrared and submillimeter emissions and utilized the universe and the redshift as a spectrograph to increase the amount of information in a collective way. From a subsample of 27 ALMA-detected galaxies at z $>$ 4.5, we built an infrared spectral energy distribution composite template. It was used to fit, with CIGALE, the 105 galaxies (detections and upper limits) in the sample from the far-ultraviolet to the far-infrared. The derived physical parameters provide information to decipher the nature of the dust cycle and of the stellar populations in these galaxies.}
  {The derived IR composite template is consistent with the galaxies in the studied sample. A delayed star formation history with $\tau_{main}$ = 500 Myrs is slightly favored by the statistical analysis as compared to a delayed with a final burst or a continuous star formation history. The position of the sample in the star formation rate (SFR) versus M$_{star}$ diagram is consistent with previous papers. The redshift evolution of the $\log M_{star}$ versus A$_{FUV}$ relation is in agreement with an evolution in redshift of this relation. This evolution is necessary to explain the cosmic evolution of the average dust attenuation of galaxies. Evolution is also observed in the L$_{dust}$ / L$_{FUV}$ (IRX) versus UV slope $\beta_{FUV}$ diagram: younger galaxies have bluer $\beta_{FUV}$. We modeled the shift of galaxies in the IRX versus the $\beta_{FUV}$ diagram with the mass-weighted age as a free parameter, and we provide an equation to make predictions. The large sample studied in this paper is generally consistent with models that assume rapid dust formation from supernovae and removal of  dust by outflows and supernovae blasts. However, we find that high mass dusty star-forming galaxies cannot be explained by the models.}
   {}
               }
   \maketitle

\section{Introduction}

Since the very first papers \cite[e.g.,][]{Madau1996} on high redshift galaxies such as the Lyman break galaxies (LBGs), the issue of how much of their energy is emitted in the far-infrared (far-IR) has been an open question in the early universe. Today, a new question is coming to the forefront and  we wonder what the dust cycle in high redshift galaxies is, that is how are large dust masses of dust formed at very high redshifts and efficiently destroyed or removed later, and what are the characteristics of these dust grains.

Thanks to deep observations with the Herschel space observatory (Herschel), the Atacama Large Millimeter/Submillimiter Array (ALMA) in the Southern Hemisphere, and the Northern Extended Millimeter Array (NOEMA) in the Northern Hemisphere, we have started to explore the dusty part of the spectral energy distributions (SEDs) of high redshift star-forming galaxies (Hiz-SFGs) \citep[][and others]{Bouwens2016, Burgarella2020, Sugahara2021, Hashimoto2019, Koprowski2020, Faisst2020a, Faisst2020b} to obtain an overall view of the energy budget of normal star-forming galaxies in the early universe.

Still, except for lensed objects, targeted studies are not very successful and most of the Hiz-SFGs are not individually detected in the rest-frame far-IR, that is in the observed submillimeter (submm) and millimeter ranges \citep[e.g.,][]{Bouwens2016, Bethermin2020, Burgarella2020, Faisst2020b}. Stacking remains the preferred tool to learn what the statistical dust properties of these galaxies are \citep[e.g., ][]{Alvarez-Marquez2016, Alvarez-Marquez2019, Carvajal2020}. 

Another interesting way to study these objects is through the far-IR fine-structure lines. These lines have been observed with the aim of understanding the interstellar medium properties and gas cooling in the neutral and ionized phases of local low-metallicity star-forming galaxies \citep[e.g., ][]{Madden2013, Cormier2019, Fernandez-Ontiveros2016}. In the high redshift universe, several of these lines have also been identified \citep[][for instance]{Harikane2014, DeBreuck2019, Cunningham2020, Pavesi2019}. We note that [OIII]88$\mu$m is very strong and may be the most intense line at high redshift where we expect that most galaxies contain a large population of O stars and have a low metallicity \citep{Arata2020}. Attempts at detecting [OIII]88$\mu$m have indeed been quite successful \citep[e.g., ][]{Inoue2016, Hashimoto2019, Alvarez-Marquez2019}. Since some of these lines (e.g., [OIII]88$\mu$m and [CII]158$\mu$m) are correlated with the star formation rate (SFR) of the galaxies, they can also be helpful when performing SED fitting by providing additional constraints on the recent SFR. Recently, the ALPINE-ALMA [CII] survey \citep{Lefevre2020, Faisst2020a, Bethermin2020} initiated an effort to detect the [CII]158$\mu$m line of 118 Hiz-SFGs in the redshift range 4.4 $<$ z $<$ 5.9. The [CII]158$\mu$m line is one of the dominant gas coolants \citep{Lagache2018}. This line is detected in the ALPINE observations with an overall detection rate of 64\% and a signal-to-noise ratio (S/N) threshold larger than 3.5$\sigma$. The ALPINE observations also allowed  for the dust continuum of 23 of the 118 galaxies above 3.5$\sigma$ to be detected \citep{Bethermin2020}. 

The dust mass (M$_{dust}$) can be estimated from the observed flux density once we can assume a dust temperature (T$_{dust}$) and other dust-related physical parameters \citep[e.g., ][]{Pozzi2021}. However, even though it might seem obvious, estimating M$_{dust}$ is much safer when using multiple data from the far-IR SEDs at different wavelengths. There is some degeneracy between T$_{dust}$ and M$_{dust}$, and more than a few data points lying both on the Rayleigh-Jeans (RJ) side, above the wavelength of the peak of the dust emission and on the Wien side, and below the peak wavelength of the dust emission which are important to constrain the SED \citep{Liang2019}. Because the individual approach is still quite a difficult task when dealing with the dust emission of galaxies close to or in the epoch of reionization (EoR), we need to combine the information from individual galaxies -- assuming they share characteristic dust emission -- to improve our wavelength coverage. 

In this paper, we adopt a statistical approach using a sample of ALMA-observed high-redshift star-forming galaxies \citep[ALPINE, ][]{Lefevre2020}  the methodology already described in  \citet{Burgarella2020} and \citet{Nanni2020}. The redshifts of the final sample cover a range 4.5 $\lesssim$ z $\lesssim$ 6.2 near or in the EoR that will be explored by the James Webb Space Telescope (JWST). The ALPINE sample is a unique source of data, which we combined to the one from \cite{Burgarella2020}\footnote{The combined ALPINE plus \cite{Burgarella2020} and \cite{Nanni2020} samples studied in this paper are referred to as our sample hereafter.} to build an IR composite template corresponding to the characteristic far-IR dust emission of Hiz-SFGs in the early universe. Some original works \citep[e.g., ][]{Ouchi1999} have tried to use local starburst galaxies to build such templates and therefore estimate upper limits of submm flux densities.  Similar to in \cite{Shapley2003} where a high S/N spectrum of z $\sim$ 3 LBGs was derived, and even in a more similar way in \cite{Pearson2013} in which 40 H-ATLAS sources with previously measured redshifts in the range 0.5 < z < 4.2 were used to derive a suitable average template for high-redshift H-ATLAS sources, the observed data from our sample were used to build the template.

In addition to using the composite IR SED to derive the dust properties of the galaxies via SED fitting, these data are also unique to help understand the stellar populations of these Hiz-SFGs and to calibrate important diagnostic diagrams in the early universe. Once an IR composite template is safely estimated from the observed ALMA detection, the spectral information in the rest-frame ultraviolet (UV), optical, and near IR is combined and we can come back to the individual approach to fit each galaxy.

In this paper, we assume a Chabrier initial mass function \citep[IMF, ][]{Chabrier2003}. We use WMAP7 cosmology \citep{Komatsu2011}.

\section{The sample of studied galaxies}

The ALPINE sample is representative of the overall Hiz-SFGs population in the redshift range 4.5 $\lesssim z \lesssim$ 5.5. It is not dominated by IR-bright (i.e., in dust continuum) submm galaxies (SMGs). The ALPINE galaxies are mainly located on or near the main sequence \citep[e.g., ][]{Rodighiero2011, Gruppioni2013, Donnari2019, Sherman2021} in the SFR versus stellar mass (M$_{star}$) relation observed at these redshifts \citep{Speagle2014, Pearson2018}. It is mostly dominated by UV-selected galaxies \citep[see Table 1 in][]{Faisst2020a} with about 62\% of the sample identified with the dropout technique, that is LBGs and 28\% that are Lyman $\alpha$ emitters which were selected with narrow bands.  We added, to the ALPINE galaxies, seven LBGs from \citet{Burgarella2020} and \cite{Nanni2020} to build our sample. The data of the latter seven LBGs were collected from several works \citep{Bouwens2016, Capak2015, Faisst2017, Scoville2016, Willott2015, Hashimoto2018}. This selection is certainly not complete and will very likely introduce biases on the results we obtained, but this is as good as we can currently do. The origins of the data are listed in Tab.~\ref{Tab.Table_Data}. 

\begin{table*}[htp]
%
\begin{tabular}{|>{\centering\arraybackslash}p{2.7cm}|>{\centering\arraybackslash}p{2.6cm}|>{\centering\arraybackslash}p{1.5cm}|>{\centering\arraybackslash}p{1.6cm}|p{8.0cm}|}
 
  \hline\hline
  {\bf Source} & {\bf Number of objects} & {\bf Selection} & {\bf redshift} &{\bf  Notes} \\
  \hline\hline
  \cite{Lefevre2020, Faisst2020a, Bethermin2020} & 118 & SFG & z $\sim$ 4.5 \newline \noindent z $\sim$ 5.5 & * 20 ALMA-7 detections with S/N $>$ 3 \newline * 78 ALMA-7 upper limits \newline * 18 ALMA-7 not selected (not enough data) \newline * S/N$_{UV-optical-NIR}$ $>$ 2.5.\newline * SFG with $>\ 5$ data points in UV-optical only \newline * [CII]158$\mu$m measurements for 64\% of the sample\\ 
\hline
  \citet{Capak2015} \& \newline \citet{Faisst2017} & 4 (HZ4, HZ6, HZ9, HZ10 & UV & z $\sim$ 5.6 & * [CII]158$\mu$m for all Hi-z LBGs \newline * ALMA-7 detections: HZ4, HZ6 (3) HZ9 \& HZ10 (5) \newline * ALMA-7 upper limits for the others \newline * HZ5 detected in Chandra and not included in the sample \newline * Additional data from \citet{Pavesi2016}\\
\hline
  \citet{Scoville2016} & 1 (566428) & UV & z = 5.89 & * ALMA-6 detection \newline * [CII]158$\mu$m measurement \\
\hline
  \citet{Willott2015} & 2 (CLM1 \& WMH5) & UV & z $\sim$ 6.1     &  * ALMA-6 detections \newline * [CII]158$\mu$m measurement \\
\hline\hline
\end{tabular}
%
  \caption{Origins of the data used in this paper. The final sample used in this paper contains 105 objects: 27 with ALMA detections and 78 with ALMA upper limits. The other objects were discarded from the sample because they do not have enough data in the UV-optical-near-IR range to perform a safe SED fitting.}\begin{center}
  \label{Tab.Table_Data}
\end{center}

\end{table*}

Using Band-7 of ALMA, the ALPINE project observed a sample of galaxies with spectroscopic redshifts in two well-observed fields: 105 galaxies in the Cosmic Evolution Survey field \citep[COSMOS,][]{Scoville2007} and the remaining 13 in the Extended Chandra Deep Field South \citep[ECDFS,][]{giacconi_vizier_2002}, with extensive multiband Hubble space telescope (HST) data from CANDELS \citep{Grogin2011, Koekemoer2011}. In this survey, 64\% of the galaxies have been detected in [CII] at $3.5\sigma$ above the noise, as well as 21\% of the galaxies detected in the continuum (S/N threshold corresponding to a 95\% purity). The sample is divided into two redshift ranges at $4.40 <  z <  4.65$ (median redshift $\langle z \rangle = 4.5$, containing 67 galaxies) and $5.05 <  z <  5.90$ (median redshift $\langle z \rangle = 5.5$, containing 51 galaxies), separated by a gap in the transmission of the atmosphere. We note that type I active galactic nuclei (AGN), identified from broad spectral lines, are excluded from the present sample.

As the galaxies included in the ALPINE sample belong to well-studied fields, they come with a rich ancillary data set \citep{Faisst2020a}. Due to the nature of the selection of these galaxies, they all have spectroscopic observations in rest-frame UV, performed with the Keck telescope and the European Very Large Telescope (VLT). A plethora of photometric observations are also available, from ground-based observatories in the UV to optical, from the Hubble space telescope (HST) in the UV, as well as from Spitzer above the Balmer break (all rest-frame features).



\section{Methodology}
\label{Sect. Methodology}

\begin{figure*}[h!]
   \centering
   \includegraphics[height=1.0\vsize]{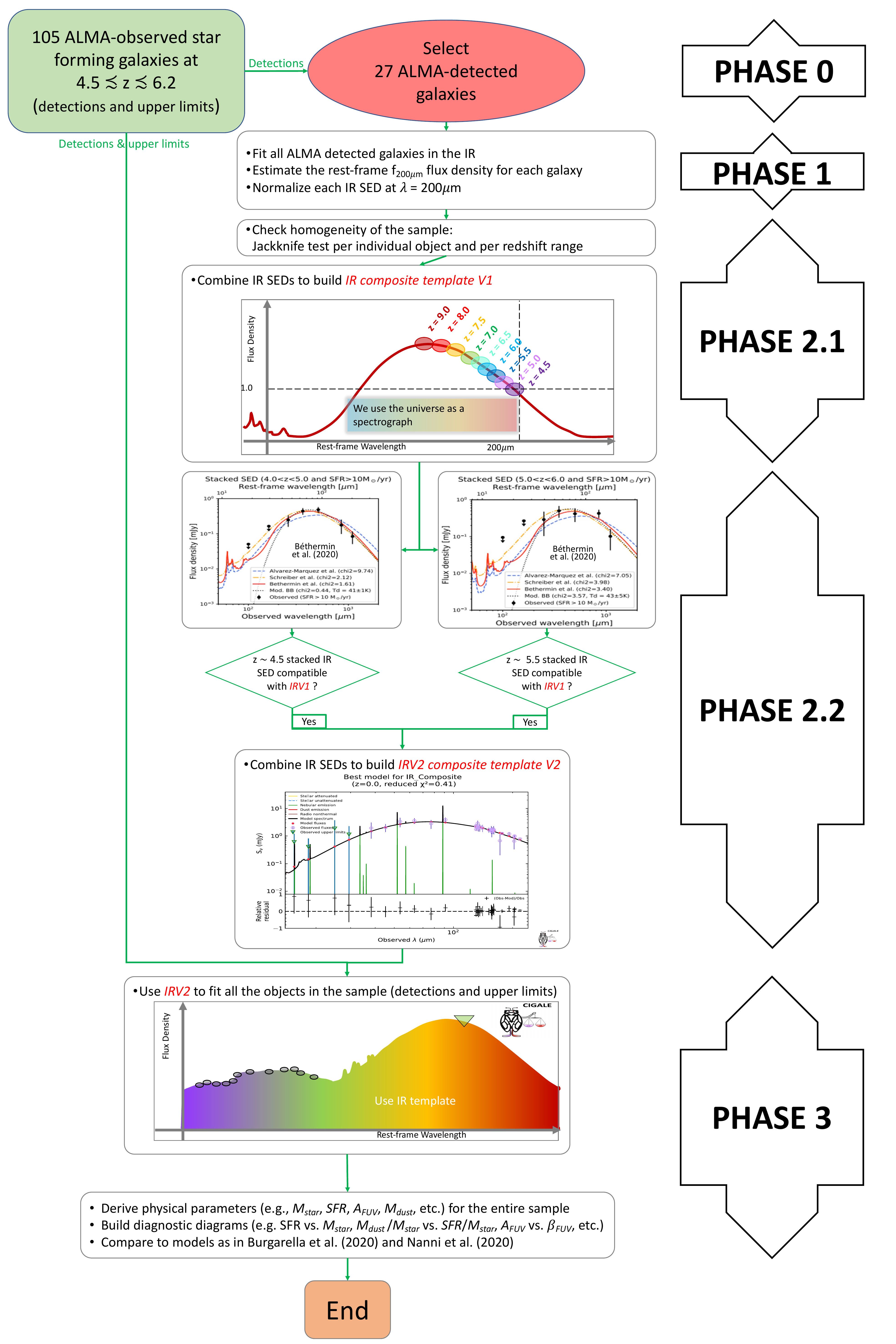}
      \caption{Flow chart of the process followed to build the composite IR SED for the Hiz-SFGs sample. The various phases of this process are shown on the right side of the figures. They are also described in more detail in the text (see Sect.~\ref{Sect. Methodology}) In the first phase, we selected only ALMA-detected objects to build the first IR template (IRV1). Next, we added the data from the z $\sim$ 4.5 and $\sim$ 5.5 stacks from \cite{Bethermin2020} to build the second and final IR template (IRV2). Adding UV and optical data to IRV2, we fit the entire (detected and upper limits) sample.}
         \label{Fig.Flowchart}
   \end{figure*}

The objective of the paper is to study the physical properties of the sample of Hiz-SFGs, and more specifically their stellar populations and dust grains. Because we do not have a wide wavelength coverage of their IR dust emission, we built an IR composite template from the subsample of ALMA-detected objects. To this aim, we applied the methodology described in Fig.~\ref{Fig.Flowchart} and it is detailed.

\subsection{Building the IR composite template} 

\subsubsection{Phase~0: Selection of the galaxies used to build the IR composite template}
From the ALPINE sample and the one compiled by \cite{Burgarella2020}, we selected the objects for which the dust continuum was detected with ALMA. More details on the selection are provided in Tab.\ref{Tab.Table_Data}. 
The final sample of ALMA-detected objects contains 27 galaxies.

\subsubsection{Phase~1: CIGALE initial individual fits of ALMA-detected galaxies to estimate the normalization at $\lambda$ = 200 $\mu$m}
The objective of this initial phase is to estimate a normalization factor for the SEDs of all the ALMA-detected galaxies. Galaxies with upper limits in the far-IR were not selected to build the IR composite SED. This normalization factor was derived from the estimated flux density at $\lambda$ = 200 $\mu$m (rest-frame), which is approximately the maximum wavelength above which we have no observed data. The SEDs of all the ALMA-detected objects at $\lambda$ = 200 $\mu$m were set to 1.0 to build the IR composite SED. All the other data were modified accordingly. 
The spectral models selected in this initial fit are described in detail in Appendix A. We selected a wide range of models, especially for the dust emission which is important for this work.

We fit all the ALMA-detected galaxies using several options with CIGALE (see Appendix~\ref{Appendix.A} for details on the priors for the models). 
CIGALE cannot model emission lines from photo-dissociation regions (PDRs). 
We used the [CII]158$\mu$m fluxes to estimate the star formation rate (SFR$_{[CII]158\mu m}$), which  we evaluated with the relation from \citet{Schaerer2020}. These SFR$_{[CII]158\mu m}$ values, along with the associated uncertainties derived from the line uncertainties, were added as properties to constrain the CIGALE SED fitting\footnote{By comparing the results of the fits with and without [CII]158$\mu$m, we get the following: M$_{star}$([w/CII]) / M$_{star}$([w/CII]) = 0.90 $\pm$ 0.27, SFR([w/CII]) / SFR([w/CII]) = 1.22 $\pm$ 0.32, L$_{dust}$([w/CII]) / L$_{dust}$([w/CII]) = 1.22 $\pm$ 0.27, and L$_{FUV}$([w/CII]) / L$_{FUV}$([w/CII]) = 1.02 $\pm$ 0.02 with no significant difference for the selected dust emission type.} To perform this analysis, we selected three types of dust emission available in CIGALE:  i) a mid-IR power law and a general modified blackbody \citep[][]{Casey2012}: PL+G\_MBB, ii) a mid-IR power law and an optically thin modified blackbody \citep{Casey2012}: PL+OT\_MBB, and, iii) the models from \citeauthor{Draine2014} \citeyearpar{Draine2014}: DL2014.

It is specifically worth checking whether the prior assumptions made on the emissivity index ($\beta_{RJ}$) of the modified blackbody could bias the final result. We could not safely estimate $\beta_{RJ}$ for the individual objects but we could do so better with the IR composite template because more data points are used, thus we also fixed $\beta_{RJ}$ = 1.0, 1.5, and 2.0 to test the impact on the level of the luminosity at 200 $\mu$m (rest-frame) used for the normalization of the individual SEDs. Table~\ref{Tab.L200um_betaRJ} shows that the bias introduced by the initial prior on $\beta_{RJ}$ is much lower that the uncertainties comings from the values derived by an SED fitting. However, we would like to stress that a wide range of SED shapes were used to estimate the normalization, including several $\beta_{RJ}$, T$_{dust}$, and DL2014 models.

\begin{table}[h!]
\newcolumntype{P}[1]{>{\centering\arraybackslash}p{#1}}
\centering

    \[$$\begin{array}{P{0.05\linewidth}|P{0.25\linewidth}|P{0.25\linewidth}|P{0.25\linewidth}}
            \hline
            \hline
            \noalign{\smallskip}
            $\beta_{RJ}$  &  $\Delta\ L_{200\mu m}=\frac{<L_{200\mu m}> - L_{200\mu m}^{\beta_{RJ}}} {<L_{200\mu m}>}$ & $\sigma(\Delta\ L_{200\mu m})$ & $[\Delta\ L_{200\mu m}]_{err}$ from  SED fitting \\                 
            \hline
            \noalign{\smallskip}
            1.0  & -0.022 & 0.118 & 0.501 \\
            1.5  & -0.013 & 0.049 & 0.466 \\
            2.0  &  0.035 & 0.060 & 0.825 \\
            \noalign{\smallskip}
            \hline
            \hline
    \end{array}$$\]
    \caption[]{Parameters of the dust emission. The results shows that whatever the prior on the value of $\beta_{RJ}$, the dispersion on the resulting normalization factor is negligible and much smaller than the uncertainties on the luminosity at $|lambda=200\mu m$ (rest frame) derived from the SED fitting and used for the normalization of the individual SEDs. This means that the method adopted to estimate the normalization factor does not significantly impact the shape of the final IR composite template.}
    \label{Tab.L200um_betaRJ}
    \end{table}

   
\subsubsection{Phase~2: Building of the IR composite template}
In this second phase, we proceeded by actually building the observed composite IR template. For this, we made use of the normalized SEDs of each of the 27 ALMA-detected objects.
   
   \begin{itemize}
       \item [$\bullet$] {\bf Phase 2.1: Checking the homogeneity in the galaxy sample $\rightarrow$ IRV1 template}: By making use of the normalization factors and benefiting from the redshift coverage (4.3 < z < 6.2), we used the universe as a spectrograph and we combined the SEDs of 27 ALMA-detected objects to form an initial version (IRV1) of the observed composite SED (Fig.~\ref{Fig.Flowchart}, Fig.~\ref{Fig.template_Ph2}). We note that this process is different from the stacking method whose aim is to detect the average flux density at a given wavelength of a sample of detected and/or undetected objects. 

   \begin{figure}[h!]
   \centering
   \includegraphics[angle=0,width=9.0cm]{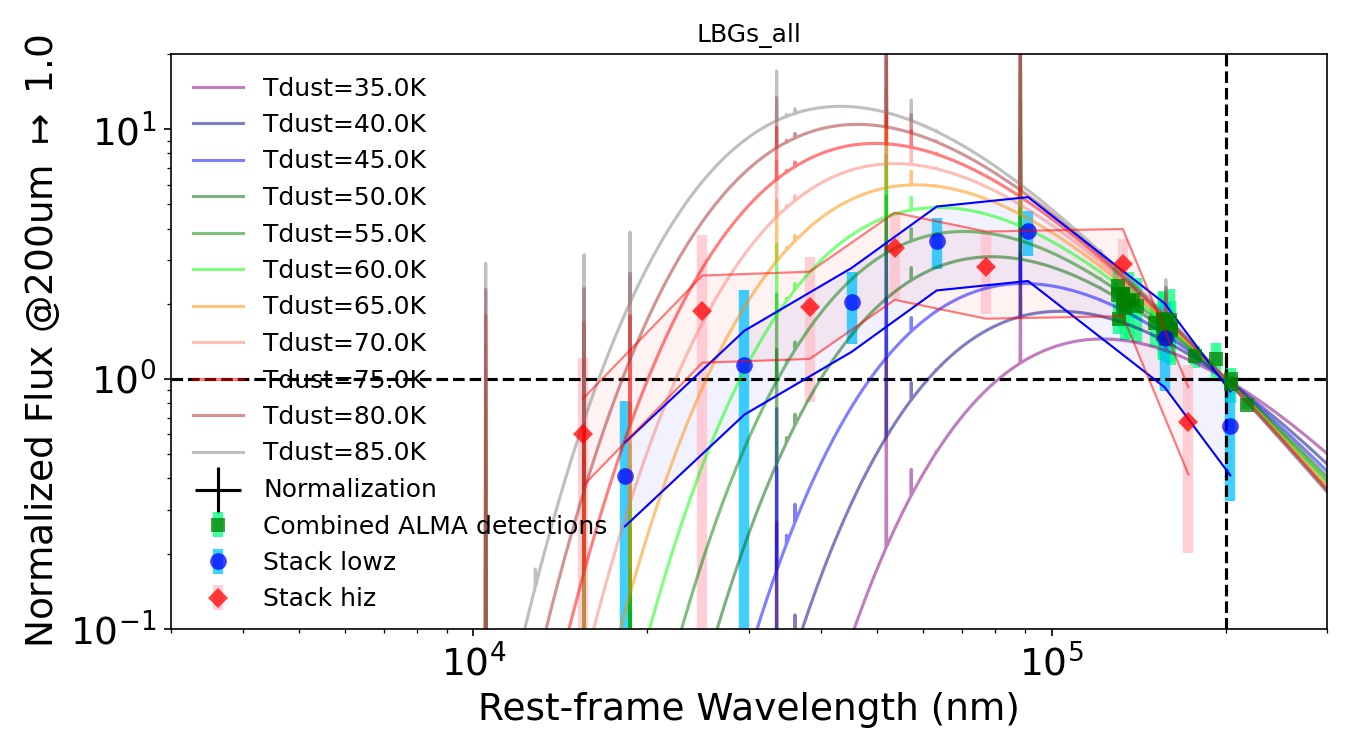}
      \caption{IR combined SED built from the ALMA-detected objects from the ALPINE sample, from \cite{Burgarella2020}, and with the two stacks from \cite{Bethermin2020} and the associated uncertainty regions. We note that the four left-most points for the stacks are upper limits, as shown by the error bars reaching the bottom of the figure. The SEDs corresponding to optically thin modified black bodies are superimposed to the observed SEDs. Qualitatively, we can see that the IR-combined SED seems to be in agreement with greenish modified blackbody SEDs, that is T$_{dust}$ $\sim$ 50 - 60K. A qualitative analysis is performed later in this paper.
              }
         \label{Fig.template_Ph2}
   \end{figure}

       We began by verifying whether the subsample of ALMA-detected galaxies is homogeneous enough to build a single IR composite template. For instance, one single object or all the objects in a given redshift range might significantly impact the parameters of the dust emission and bias the IR composite template. Here, we only show the results of this test assuming PL+OT\_MBB, but we checked that the results were the same regardless of the assumed dust emission option. 
       
       To reach this goal, we used the knife-jacking method, where we removed the SED of each galaxy (or a series of galaxies in a redshift range) sequentially from the list of objects, built the composite template, and fit it with CIGALE. Then, we put the removed objects back in the detected subsample and we reproduced this operation recursively as many times as there were objects (or series of galaxies in a redshift range) in the subsample. At the end, we could compare the parameters from the several composite templates built from $N_{obj}-N_{removed-obj}$ and we checked whether they were consistent. 
       
       From the analysis on individual objects (Fig.~\ref{Fig.jcknf_Tdust_betaRJ}), we discovered that only when removing HZ10 at z=5.659 did we get slightly different values for T$_{dust}$ and $\beta_{RJ}$.\ However, it does not significantly bias the parameters defining the shape of the composite template. We decided to keep it in the sample because it contributes to the representativity of the studied galaxies. The dust parameters are very stable with average values at T$_{dust}$ $\approx$ 54K and $\beta_{RJ}$ $\approx$ 0.9. 

   \begin{figure}[h!]
   \centering
   \includegraphics[angle=0,width=9cm]{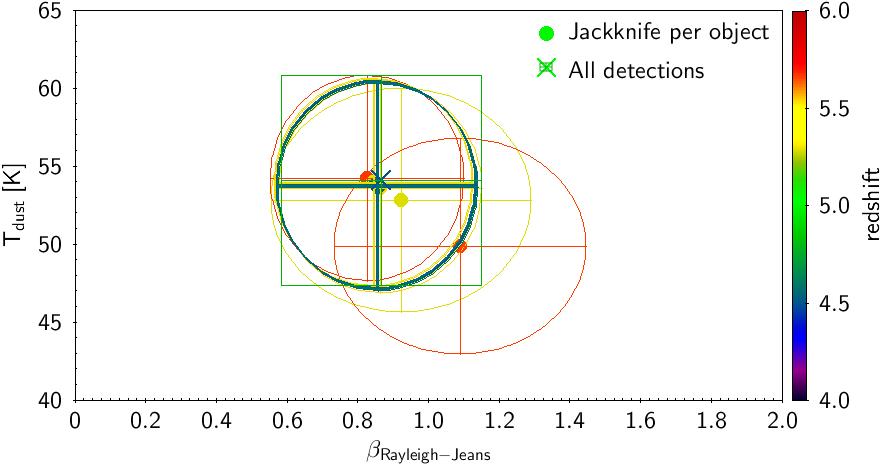}
      \caption{Degeneracies in T$_{dust}$ and $\beta_{RJ}$. The knife jacking method allowed us to estimate whether the characteristics of the dust emission, i.e., T$_{dust}$ and $\beta_{RJ}$, vary when removing and adding back individual galaxies from the sample used to build the composite template. The ellipses show the uncertainties. The outlier point at $\beta_{RJ}\sim1.1$ and T$_{dust}\sim51$K was obtained when taking off HZ10 at z=5.659 with five pieces of data in the far-IR+submm range. Only when removing it did the average locus move to slightly larger values of $\beta_{RJ}$, but with larger uncertainties. }
      \label{Fig.jcknf_Tdust_betaRJ}
   \end{figure}

       From the analysis on redshift ranges, we discovered that the redshift range 5.0 $<$ z $<$ 6.0 brings the objects with the widest wavelength range to the IR composite SED. Without the ALMA-detected galaxies in the range 5.0 $<$ z $<$ 6.0, there was a difference for T$_{dust}$ and $\beta_{RJ}$ (Tab. \ref{Tab.jcknf_Tdust_betaRJ}). But again, the results are in agreement within the uncertainties, even though an evolution within the uncertainties cannot be ruled out.
       
     This second phase provided an IR composite template IRV1 shown as green points in Fig.~\ref{Fig.template_Ph2}.

\begin{table}[h!]
\centering
    
    \[$$\begin{array}{p{0.26\linewidth}p{0.20\linewidth}p{0.19\linewidth}p{0.19\linewidth}}
            \hline
            \hline
            \noalign{\smallskip}
            removed z range  &  N$_{removed-obj}$ & $\beta_{RJ}$ & T$_{dust}$ \\                 
            \hline
            \noalign{\smallskip}
            4.0 $<$ z $<$ 5.0 & 62 & 0.84 $\pm$ 0.28 & 54.3 $\pm$ 6.7 \\
            5.0 $<$ z $<$ 6.0 & 41 & 1.30 $\pm$ 0.55 & 47.6 $\pm$ 8.4 \\
            6.0 $<$ z $<$ 7.0 &  2 & 0.85 $\pm$ 0.28 & 53.7 $\pm$ 6.6 \\
            \noalign{\smallskip}
            \hline
            \hline
    \end{array}$$\]
    \caption[]{Degeneracy of $\beta_{RJ}$ and T$_{dust}$. Results of the knife jacking method applied on the entire sample when removing and adding back objects in the given redshift ranges from the first column to build the composite template. The dust parameters are stable. Even though it is still consistent with the other ones, the results without one redshift range (namely 5.0 $<$ z $<$ 6.0) differ. This could be explained because the 5.0 $<$ z $<$ 6.0 range contains the widest wavelength range and it is therefore more constraining to estimate $\beta_{RJ}$ and T$_{dust}$. So, we cannot rule out a small variation of these two parameters with the redshift inside the quoted uncertainties.} \centering
    \label{Tab.jcknf_Tdust_betaRJ}
    \end{table}
      
       \item [$\bullet$] {\bf Phase 2.2: Checking if the stacked data from \cite{Bethermin2020} are in agreement with the IRV1 $\rightarrow$ IRV2 template}: In addition to the main ALPINE data, \cite{Bethermin2020} stacked data from two galaxy samples at z $\sim$ 4.5 and z $\sim$ 5.5 with Herschel \citep{Pilbratt2010} data from the PEP \citep{Lutz2011} and HerMES \citep{Oliver2012} surveys and AzTEC/ASTE data from \cite{Aretxaga2011} at 1.1 mm. At 850 µm, they used the SCUBA2 data from \cite{Casey2013}. This data set can be very useful because it extends the wavelength range to the mid-IR. However, because there are not enough ALPINE sources to obtain a sufficiently high S/N in the stacked Herschel data, the ALPINE selection was not used for the stacking. \cite{Bethermin2020} define two redshift bins (4 $<$ z $<$ 5 and 5 $<$ z $<$ 6) that match the redshift ranges probed by ALPINE. 
       Before further continuing to fulfill our task of building the LBG composite IR template, we first need to assess whether the z $\sim$ 4.5 and/or z $\sim$ 5.5 stacks are consistent with the IRV1 composite IR template.
       
       Our analysis allows us to conclude that both stacks are consistent (all $\chi^2_\nu$ from the SED fitting have $\sim 0.3 - 0.4$) with the ALMA-detected galaxies' composite IR template built by combining the full sample of the 27 ALMA-detected objects. The final IR composite template (IRV2, Fig.~\ref{Fig.template_Ph2}) was built from the ALMA-detected galaxies from \cite{Burgarella2020}, the ALPINE sample, in addition to the two stacks from \cite{Bethermin2020}. 
       
       
       In Tab.~\ref{Tab.SED_IR.composite.data}, we provide the IR composite template based on the observed data and show the fits with the three dust emissions from 1 $\mu m$ to 1000 $\mu m$ fitted in Fig.~\ref{Fig.template_Ph1.1}. The modeled SEDs from 20 $\mu$m to 1 mm are listed in Tab.~\ref{Tab.SED_IR.composite.models}.
   \end{itemize}

   \begin{figure}[!]
   \centering
   \includegraphics[angle=0,width=9.2cm]{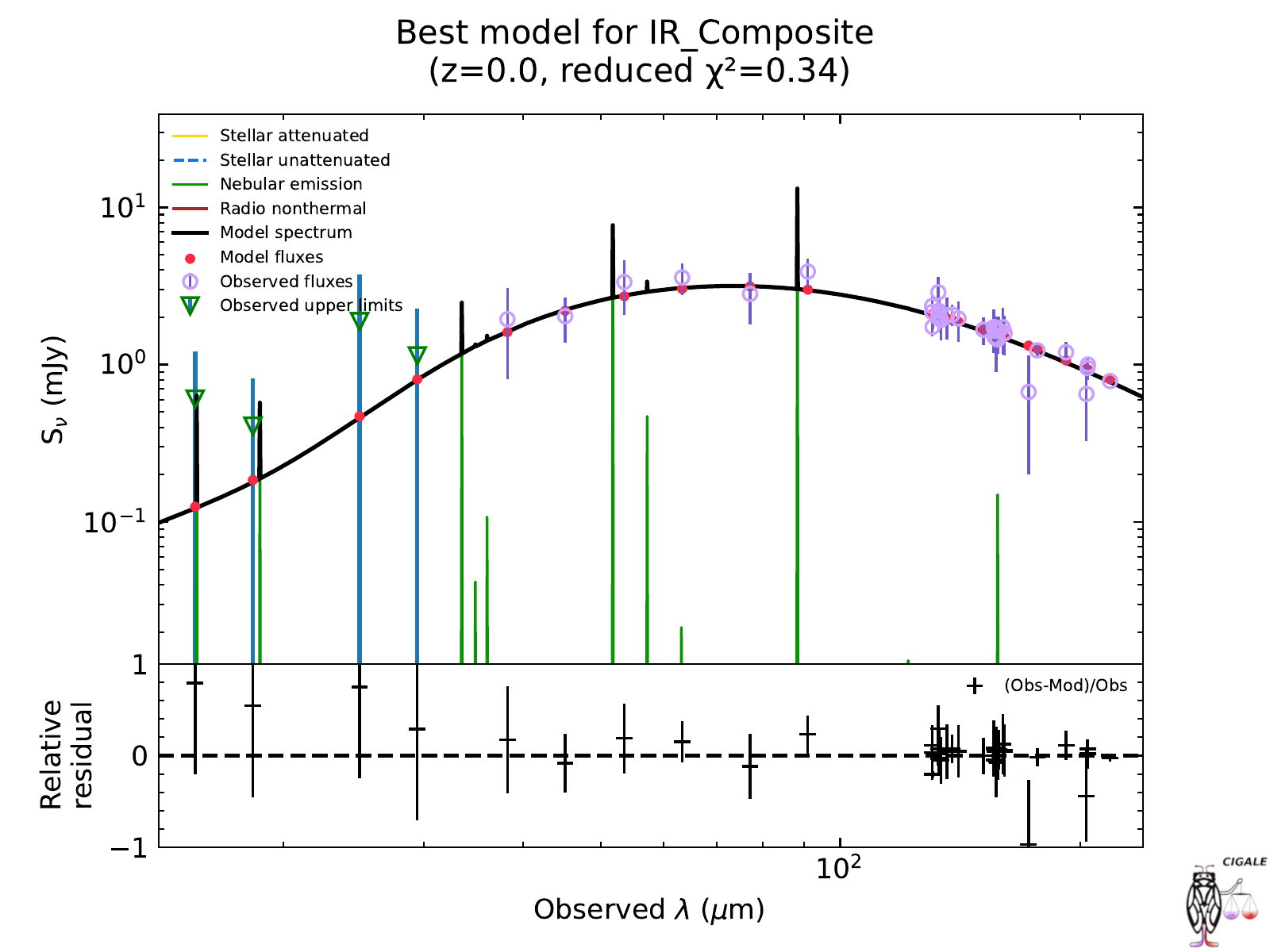}
   \includegraphics[angle=0,width=9.2cm]{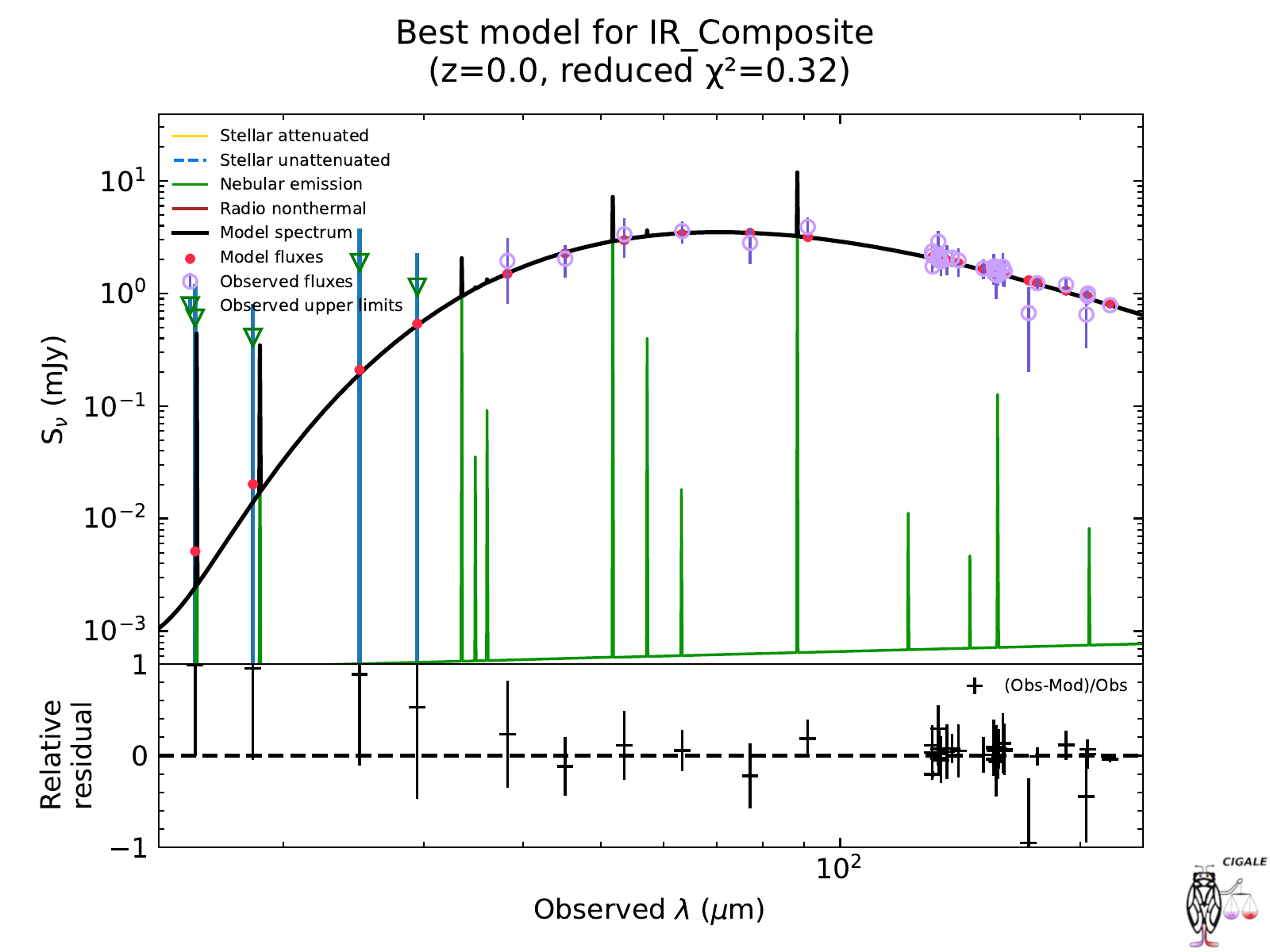}
   \includegraphics[angle=0,width=9.2cm]{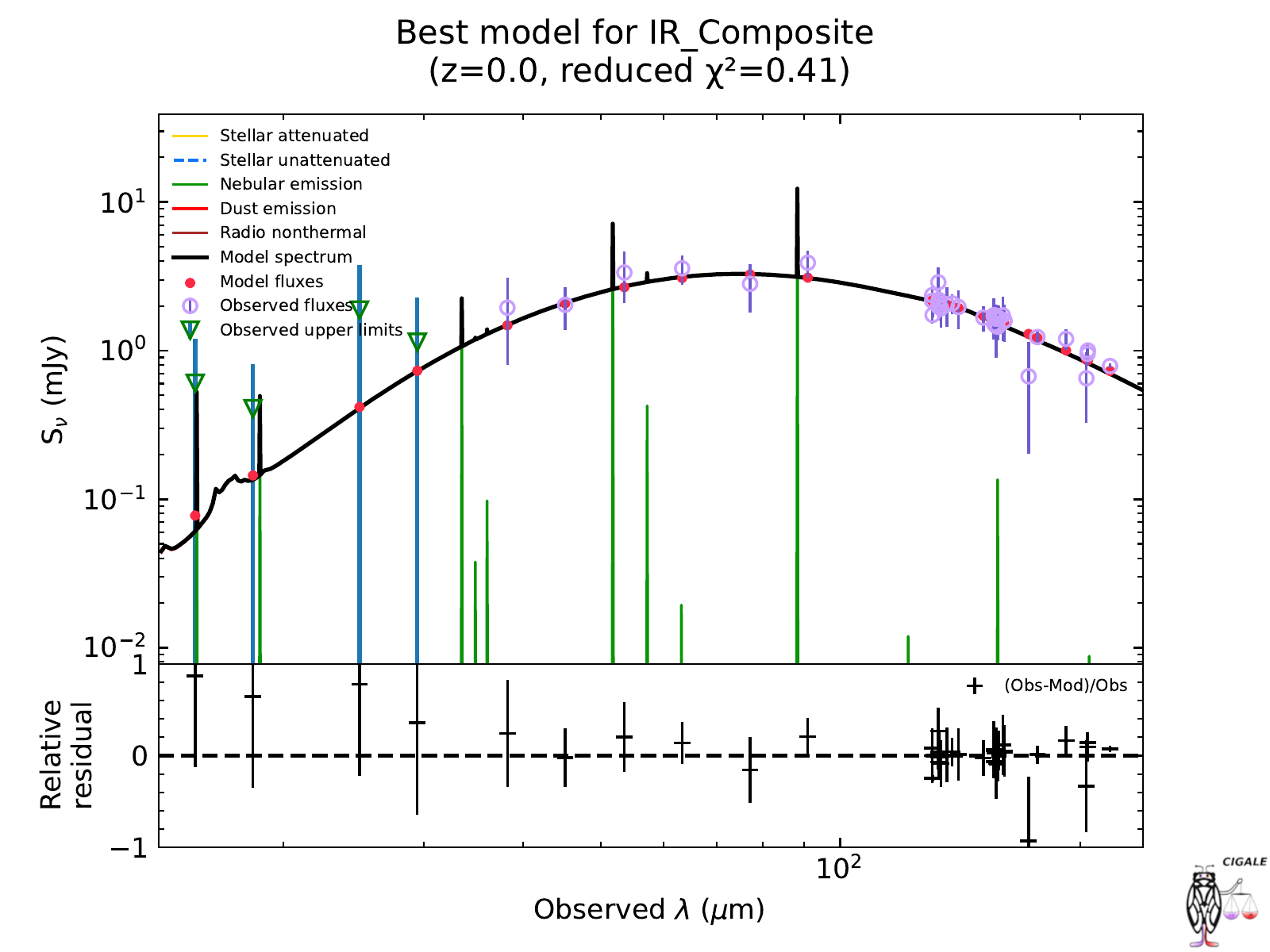}
      \caption{Comparison of the various fits of the IR composite SED built with data from \cite{Burgarella2020}, from ALPINE, and from the z $\sim$ 4.5 and z $\sim$ 5.5 stacks from \cite {Bethermin2020}. Top:  Fit with a modified general blackbody and power law in the mid-IR as in Casey et al. (2012). Middle: Fit with an optically thin modified blackbody and power law in the mid-IR as in Casey et al. (2012). Bottom: Fit with a model from Draine \& Li (2014). The fits are all statistically equivalently good with reduced $\chi^2$ $\approx$ 0.3 - 0.4.}
      \label{Fig.template_Ph1.1}
   \end{figure}


\subsubsection{Fitting the IR composite template}
The main dust parameters derived from fitting the observed IRV2 IR composite template (Fig.~\ref{Fig.template_Ph1.1}) are listed in Tab.~\ref{Tab.fit_IR_template}. The objects in the present sample are Hiz-SFGs with a low dust attenuation \citep[e.g.,][ and later in this paper]{Faisst2020a}. This assumption is valid for the ALPINE sample \citep{Faisst2020a}. Therefore, we make the assumption that the best emission models for these objects should be optically thin. We checked this hypothesis by estimating the following \citep[Eq.~2 in ][]{Jones2020}: 

$$\tau_\nu = \frac{M_{dust}}{A_{gal}}\kappa_{\nu}$$

where $\tau_\nu$ is the optical depth, M$_{dust}$ is the dust mass, A$_{gal}$ is the area covered by the galaxy, and $\kappa_{\nu}$ is the dust mass absorption coefficient: $$\kappa_{\nu}\ =\ \kappa_0\ (\nu\ / \nu_0)\ ^{\beta_{RJ}}$$ where $\nu_0$ is the frequency where the optical depth equals unity and $\beta_{RJ}$ is the spectral emissivity index from the Rayleigh-Jeans range. We tested the opacity both with the optically thin and with the general modified black bodies \citep{Casey2012}. The median radius of the galaxies are $r_{e,[CII]}$ = 2.1 $\pm$ 0.16 kpc \citep{Fujimoto2020} to estimate $A_{gal}$. Both provide values that are in the range [10$^{-4}$ - 10$^{-2}$], that is $\tau_\nu \ll 1.0$, which confirms that the optically thin assumption is valid here. In the rest of the paper, we only make use of DL2014 and PL+OT\_MBB dust emissions. 

\begin{figure}
   \centering
   \includegraphics[angle=0,width=9cm]{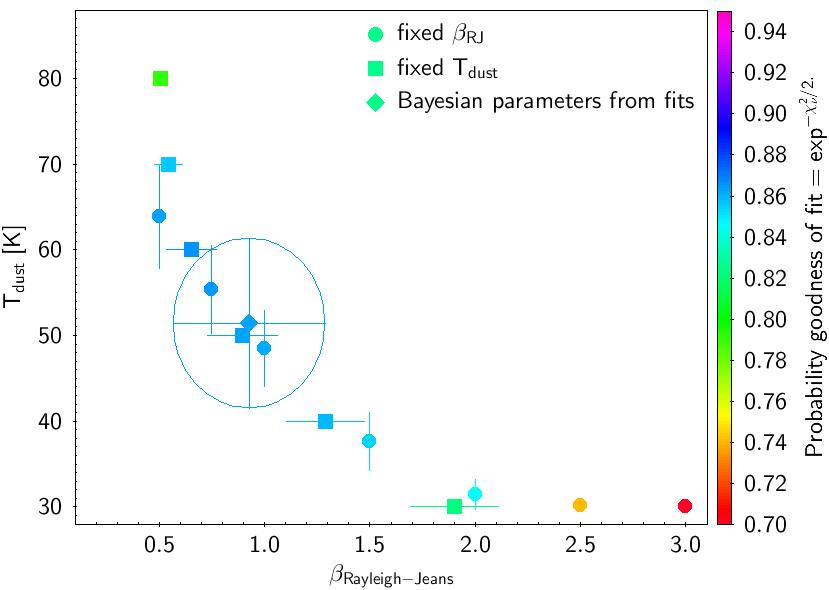}
      \caption{Results of the tests on the determination of T$_{dust}$ and $\beta_{RJ}$ and their possible degeneracy are shown here. Dots show the evolution of T$_{dust}$ when $\beta_{RJ}$ was fixed during the fits. Boxes show the evolution of $\beta_{RJ}$ when T$_{dust}$ was fixed during the fits. Finally, the diamond symbol shows the result and uncertainties on both axes when T$_{dust}$ and $\beta_{RJ}$ are both free. The color coding indicates the level of the goodness of fit. 
      }
      \label{Fig.degeneracy_Tdust_beta}
   \end{figure}

After fitting the IR composite template, we derived values for T$_{dust}$ and $\beta_{RJ}$ (Tab.~\ref{Tab.fit_IR_template}). However, these dust parameters are known to be degenerate depending on the S/N and the wavelength sampling of the far-IR SED \citep[e.g.,][]{Juvela2013, Tabatabaei2014}. This degeneracy is a serious problem when using only one or a few data points in the far-IR. However, using the derived IR composite template helps to remove the degeneracy. We tested how well T$_{dust}$ and $\beta_{RJ}$ can be estimated by performing fits with fixed T$_{dust}$ and varying $\beta_{RJ}$, then by fixing $\beta_{RJ}$ and keeping T$_{dust}$ free. The results are compared to the parameters derived by keeping both parameters free in Fig.~\ref{Fig.degeneracy_Tdust_beta}. Even though we can see the usual regular evolution of T$_{dust}$ with  $\beta_{RJ}$ (or vice versa), the quality of the fit with fixed parameters improves when getting closer to the Bayesian values derived with CIGALE when both parameters are free (Tab.~\ref{Tab.fit_IR_template}). 

\begin{table}[h!]
    \[$$\begin{array}{p{0.25\linewidth}p{0.20\linewidth}p{0.20\linewidth}p{0.20\linewidth}}
            \hline
            \hline
            \noalign{\smallskip}
              &  PL+G\_MBB & PL+OT\_MBB & DL2014 \\                 
            \hline
            \noalign{\smallskip}
            $\alpha_{MIR}$    &  2.23$\pm$0.63  & 2.00$\pm$0.82 & N/A \\
            $\beta_{RJ}$     &  1.43$\pm$0.47  & 0.87$\pm$0.28 & N/A \\
            T$_{dust}$[K]  &  65.5$\pm$5.1 & 54.1$\pm$6.7  & N/A \\
                        \hline
q$_{PAH}$      &  N/A & N/A & 0.47 \\
            $\alpha$      &  N/A & N/A & 2.39$\pm$0.44 \\
            u$_{min}$     &  N/A & N/A & 18.1$\pm$12.7 \\
            $\gamma$      &  N/A & N/A & 0.54$\pm$0.35 \\
            \hline            \noalign{\smallskip}
            L$_{dust}$/$10^{20}$ [W]  &  2.84$\pm$0.14  & 2.43$\pm$0.12 & 2.57$\pm$0.13 \\
            \hline            \noalign{\smallskip}
            \hline
            \hline
    \end{array}$$\]
    \caption[]{Main relevant physical parameters derived by fitting the IR template with the various assumptions of the dust emission: PL+G\_MBB is based on a power law in the mid-IR plus the general blackbody formula as in Casey et al. (2012); PL+OT\_MBB is a power law in the mid-IR ($\alpha_{MIR}$) plus an optically thin blackbody, which is again similar to Casey et al. (2012); and DL2014 stands for Draine \& Li (2014) models. The parameters are described in Appendix A. Because the SEDs were normalized to 1.0 at $\lambda =200\mu m$, the values of L$_{dust}$ can be directly compared.}
    \label{Tab.fit_IR_template}
\end{table}

      The value of $\beta_{RJ}$ = 0.87 $\pm$ 0.28 for a power law plus optically thin modified blackbody found in this paper is low, but comparable to the minimum values found by \citet[][]{Bendo2003, Galametz2012}, for example, for nearby galaxies in the range $0.8 < \beta_{RJ} < 2.5$. Lower $\beta_{RJ}$ closer to 1.0 were estimated when fitting different types of galaxies and more specifically very low-metallicity galaxies such as SBS~0335-052 \citep{Hunt2014}, NGC~1705 \citep{OHalloran2010}, and even low-metallicity regions in very nearby galaxies with excellent coverage to estimate $\beta_{RJ}$ similar to Messier 33, for instance \citep{Tibbs2018, Tabatabaei2014}. \citet{Tabatabaei2014} found an apparent decrease in both T$_{dust}$ and $\beta_{RJ}$ with an increasing M33 radius. This corresponds to regions with low metallicities and they propose that the effect could help to find an origin in the different grain compositions and, possibly, different size distributions. Assuming such $\beta_{RJ}$ values for Hiz-SFGs would mean that the mean dust temperature could be higher than T$_{dust}$ estimated with emissivities in the range 1.5 $< \beta_{RJ} <$ 2.0. This result ($\beta_{RJ} \sim 1$) needs to be further confirmed for the SEDs of similar low-M$_{star}$, low-A$_{FUV}$ galaxies with a better RJ wavelength coverage.
      
     The value of T$_{dust}$ found from the IR composite template (54.1 $\pm$ 6.7K) should be compared to other ALMA-based dust temperatures for objects at high redshift (z $>$ 4.5). 
     In a recent paper, \citet{Bakx2021} present the evolution of T$_{dust}$ for ''normal'' (i.e., main sequence) galaxies from z $\sim$ 0 to z $\sim$ 8. A linear increase is suggested up to z = 6. At larger redshifts, a large dispersion in T$_{dust}$ can be noticed even though the linear relation might hold, given the large uncertainties \citep{Faisst2020b, Harikane2020, Sugahara2021, Laporte2019}. \cite{Bakx2021} show that adding ALMA Band 9 significantly reduces the uncertainty on the dust temperature for a single object by further constraining the shape of the SED. 
     At the mean redshift z = 4.94 $\pm$ 0.54 of our entire sample, the dust temperature that could be estimated assuming the linear relation from Fig. 4 in \cite{Bakx2021} would be T$_{dust}$ $\approx$ 49K, which is not significantly different from our estimation of 54.1 $\pm$ 6.7K. However, \cite{Bakx2021} derived a value of $\beta_{RJ}$ = 1.61 $\pm$ 0.60 which is larger than the one found in the previous paragraph. Both are consistent if we account for the uncertainties. Lower values for $\beta_{RJ}$ mean higher dust temperatures, as shown in Fig.~\ref{Fig.degeneracy_Tdust_beta}. Thus measuring $\beta_{RJ}$ with a good sampling of the RJ part of the spectrum is important to address the question about the decrease in $\beta_{RJ}$ in low-metallicity regions toward the outskirts of local galaxies and in some local low-metallicity galaxies. If so, the increase in dust temperature might be larger than when evaluated with $\beta_{RJ}$ $\sim$ 1.5 - 2.0.
      

\subsection{Phase 3: Far-UV to far-IR fit of individual galaxies with the IR composite template} 
We now fit the complete sample of objects 
assuming the dust emission from the IRV2 composite SED. 
The rest-frame UV, optical, and near IR ranges are useful when constraining the stellar emission and, therefore, the properties of the stellar populations. However, attenuation also impacts the rest-frame UV spectral range, which is therefore useful to constrain the amount of dust attenuation. The assumptions used when fitting the entire galaxy sample with CIGALE are given in Tab.~\ref{Tab.CIGALE.Ph2}. We focus on the properties of the stellar populations and the dust attenuation in these objects. 

Even though the Bayes factor is conceptually more conclusive, its computation can still be very complex. We need to compute a quantity called the marginal likelihood or evidence ($P(D/M_i)$), where $D$ are the photometric data and $M_i$ are each model, which implies computing a very complicated and time-consuming  integral because the likelihood of observation $D$ under the model $M_i$ must quantify over all possible parameters of that model. An analytic computation is almost never possible and direct evaluation by numerical quadrature is almost never feasible for models of real-world dimensionality and complexity. This is why a variety of approximations based on special properties of the models or their posterior distributions were developed \cite{Kass1995}.

Generally, it is found that the conclusions drawn from the Bayes factor  are more satisfying, but also more complex methodologies have not been qualitatively very different from those drawn from the simpler BIC method \citep{Raftery1998}. However, we can elaborate on the use of the BIC versus Bayes Factor for our specific case. From \cite{Raftery1998}, we determined that in our case, the unit information prior should reasonably cover the range of observed data because we have a homogeneous sample of galaxies and we have a rough idea of the general range within which the data are likely to lie in advance. So we defined our priors to homogeneously cover the expected range of parameters. \cite{Volinsky1997} showed, via simulations, that the performance of Bayes factors can be better than that of BIC if the prior used is spread out less than the unit information prior. In our case, the unit information prior is quite spread out (see Table 5) and BIC should be at least as good as the Bayes factor.  Because of this spread out prior, BIC provides more conservative results.\ Also, if an effect is favored by BIC, this should be a solid result. On the other hand, if BIC does not favor an effect, it might still be possible that another, justifiable prior could change the conclusion. For us, that means that the most conclusive results would be that an SFH that includes a burst is solidly ruled out when compared to a delayed SFH with $\tau$ = 500 Myrs SFH.

\begin{figure}[h!]
   \centering
       \includegraphics[angle=0,width=9cm, height=5.5cm]{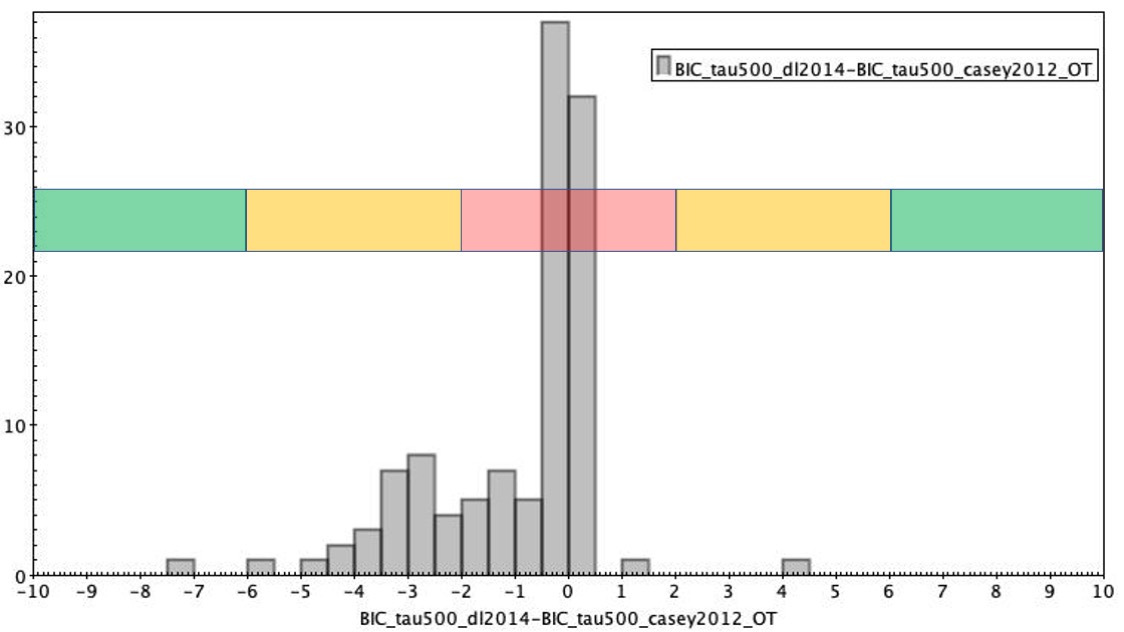}
       \includegraphics[angle=0,width=9cm, height=5.5cm]{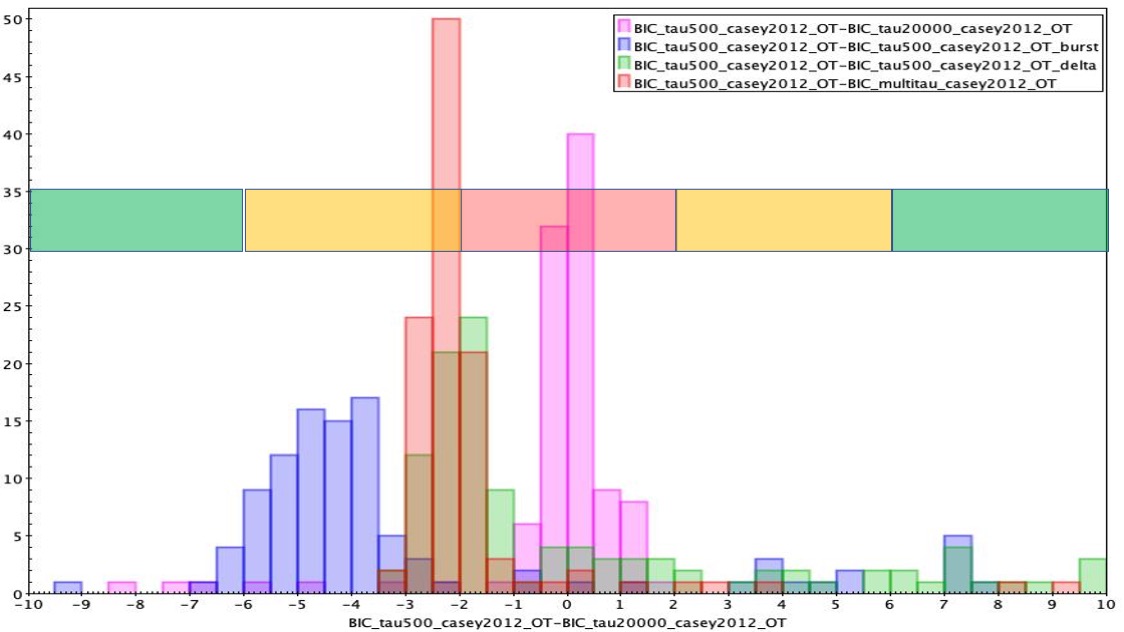}
       \includegraphics[angle=0,width=9cm, height=5.5cm]{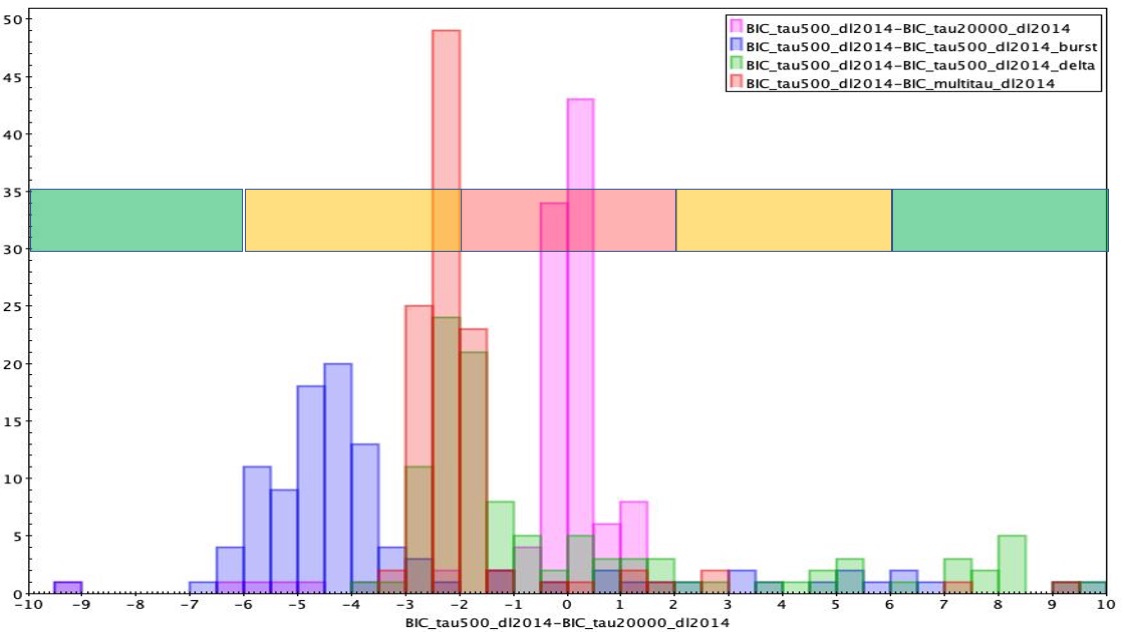}

      \caption{Result of the $\Delta$BIC test on our sample of Hiz-SFGs. A delayed SFH without a burst and $\tau_{main}$ = 500 Myrs is labeled as ”tau500”. With an additional burst at the end of the SFH, it is labeled as ”burst”. A constant SFH without a burst and $\tau_{main}$ = 20 Gyrs is labeled as "tau20000". And when several $\tau_{main}$ could be selected in the SED fitting, it is labeled as ”multitau” in the legend. Top: $\Delta$ BIC test that compares the influence of the DL2014 model and the PL+OT\_MBB dust emissions in building the IR template. Center: $\Delta$BIC test on the SFH assuming the PL+OT\_MBB for the IR template. Bottom: $\Delta$BIC test on the SFH assuming the DL2014 model for the IR template. The color band allows one to interpret the results of the evidence ($\Delta$BIC) against the model with the higher BIC: red means "faint evidence," orange means "positive evidence," and green mean "strong evidence." We do not see any strong evidence that DL2014 or PL+OT\_MBB are better for fitting the data. An SFH that includes a burst is positively ruled out, while a delayed SFH with $\tau$ = 500 Myrs is weekly favored.
              }
         \label{Fig.BIC_results}
   \end{figure}

We decided to use the Bayesian information criterion (BIC) test to compare the various hypotheses made on the dust emission and on the star formation history in Fig.~\ref{Fig.BIC_results}. To interpret the results from the BIC tests, we refer to \citet{Kass1995}: $$BIC\ = \ \chi ^{2}\ +\ k\ln(n)$$ where $k$ is the number of model parameters in the test and $n$ is the sample size, that is to say the number of pieces of photometric data in our case. When comparing several models, the one with the lowest BIC is preferred. To interpret the results, $ 0 < \Delta BIC < 2$ means that evidence for a difference between the two hypotheses is faint ($ 2 < \Delta BIC < 6$ means positive, $ 6 < \Delta BIC < 10$ means strong, and $\Delta BIC > 10$ means very strong).

We start by comparing the two dust emission models, DL2014 \citep{Draine2014} and PL+OT\_MBB \citep{Casey2012}, available in CIGALE. Tab.~\ref{Tab.BIC_results} presents all the results of the BIC analysis on our sample of Hiz-SFGs, numerically. The histogram of $\Delta$BIC values strongly peaks at 0 with a small tail extending to $\Delta$BIC = -4 (Fig.~\ref{Fig.BIC_results}), meaning that we do not see any strong difference between DL2014 and PL+OT\_MBB when fitting our sample. However, for a minority of objects, PL+OT\_MBB might be favored. We show the results using both DL2014 and PL+OT\_MBB later in the paper. However, in general, no significant difference as to the quality of the fit is observed, confirming the BIC analysis.

   \begin{table*}[h]
       \centering
          \resizebox{1\columnwidth}{!}{\begin{tabular}{cccccccccc}
          \hline \hline \\ 
          \multicolumn{9}{c}{DL2014 vs. PL+OT\_MBB}\\
          \hline 
          $\tau_{main}$[Myr] & k & Age$_{main}$[Myrs] & $\tau_{burst}$[Myr] & Age$_{burst}$[Myrs] & $\delta$ & Mean & SD & Min & Max \\
          \hline
          500 Myrs & 2 $\leq k \leq 5$ & 101 values in [2, 1200] & & & & -0.95 & 1.58 & -7.28 & 4.08  \\  
          \hline \hline \\ 
          \multicolumn{9}{c}{DL2014}\\
          \hline 
          $\tau_{main}$[Myr] & k & Age$_{main}$[Myrs] & $\tau_{burst}$[Myr] & Age$_{burst}$[Myrs] & $\delta$ & Mean \\
          \hline
          20000 Myrs & 2 $\leq k \leq 5$ & 101 values in [2, 1200] & & & & -3.35 & 11.21 & -68.37 & 1.39 \\
          10 values in [100, 1000] & 3 $\leq k \leq 6$ & 101 values in [2, 1200] & & & & -0.75 & 5.83 & -3.15 & 41.64  \\
          500 & 3 $\leq k \leq 6$ & 101 values in [100, 1200] & 20000 & 2, 5, 10, 20, 50 & & -1.62 & 5.87 & -9.23 & 16.89  \\
          500 & 3 $\leq k \leq 6$ & 101 values in [2, 1200] & & & 7 values in [-0.70, 0.70] & 1.49 & 7.12 & -3.59 & 43.43  \\
         \hline \\
          \multicolumn{9}{c}{PL+OT\_MBB}\\
          \hline
          20000 Myrs & 2 $\leq k \leq 5$ & 101 values in [2, 1200] & & & & -3.49 & 11.40 & -68.37 & 1.55  \\
          10 values in [100, 1000] & 3 $\leq k \leq 6$ & 101 values in [2, 1200] & & & & -0.67 & 6.00 & -3.15 & 41.64  \\
          500 & 3 $\leq k \leq 6$ & 101 values in [100, 1200] & 20000 & 2, 5, 10, 20, 50 & & -1.05 & 6.57 & -9.23 & 19.16  \\
          500 & 3 $\leq k \leq 6$ & 101 values in [2, 1200] & & & 7 values in [-0.70, 0.70] & 1.51 & 7.20 & -3.04 & 42.94  \\
          \hline \hline
      \end{tabular}}
       \caption{Comparison of the reference model (SFH delayed: $t / \tau^2\ \exp(-t/\tau)$ with $\tau$ = 500 Myrs) to all the others using the BIC. This comparison was performed by fitting all the objects in our sample. We note that the IR composite SED is not used in this analysis, but only individual ALMA measurements (detections and upper limits). We first assumed a DL2014 model for the IR template. The, we assumed a PL+OT\_MBB for the IR template. The BIC values do not change much for the two options for the IR template. The range in the number of parameters $k$ is also indicated for each series of run. To the parameters listed in the table, we need to add the amount of dust attenuation, i.e., E\_BVs\_young. The sample has at least five photometric points in the rest-frame UV and optical as well as one photometric point in the rest-frame far-IR, so $n \geq 7$. More precisely, the complete sample of individual galaxies have $7 \leq n \leq 24$ with <n> = 13.3 $\pm$ 4.3 and 11/119 objects have more than ten photometric points. }\label{Tab.BIC_results}
    \end{table*}

We now focus on the comparison between the various SFHs. Fig.~\ref{Fig.BIC_results} shows the results of the BIC analysis by comparing the various kinds of SFHs used in the CIGALE analysis. 
We note that we only used the initial observed data and not the IR composite template. The only significant difference is that a simple delayed SFH with $\tau_{main}$ = 500 Myrs without any final burst is preferred to a delayed SFH with a burst. A constant SFH (with $\tau_{main}$ = 20 Gyrs) and multi-$\tau_{main}$ are weakly disfavored when compared to a simple delayed SFH with $\tau_{main}$ = 500 Myrs. These comparisons are compatible with both DL2014 and PL+OT\_MBB. There is some dispersion in the results which suggests that the preferred modeling for the entire galaxy population might not be valid for each individual galaxy, as shown in Fig.~\ref{Fig.BIC_results}. 


    In order to move a little bit further with our comparison of the models, we performed a mock analysis with CIGALE from the initial fit, again without using the IR composite template. We selected the PL+OT\_MBB and DL2014 options for the dust emission in Fig.~\ref{Fig.Mocks}. In brief, for each object, the CIGALE mock analysis consists in using the best fit model for each object to generate an input photometric catalog (called a mock catalog) to which we randomly added uncertainties drawn inside a Gaussian distribution with $\sigma$ corresponding to the observed uncertainties \citep[see][]{Boquien2019}. The SED fitting procedure was applied in the very same way as when fitting the observed SEDs. Then, we compared the derived output parameters to those of the known input (Fig.~\ref{Fig.Mocks}). The results from this mock analysis (Tab.~\ref{Tab.Mocks}) show that both for the PL+OT\_MBB and for the DL2014 dust emissions, we were able to recover the main physical parameters related to the SFH and to the dust emission. This means that we do not expect strong degeneracies ($r^2\ = \ 0.80$ shown in Fig.~\ref{Fig.Mocks}, where $r$ is Pearson’s correlation coefficient commonly used in linear regression). The mean value and standard dispersion of the main parameters derived from the individual fits of our galaxy sample are given in Tab.~\ref{Tab.Results_CIGALE.Ph3} and the individual values for the same parameters are listed in Tab.~\ref{Tab.Physical_parameters_dl2014} and Tab.~\ref{Tab.Physical_parameters_dl2014} in Appendix E (available online) in their entirety.

   \begin{table*}
      \centering
       \resizebox{2\columnwidth}{!}{
          \begin{tabular}{cccccccccccccc}
          \hline
            \noalign{\smallskip}
            Run & L$_{dust}$ & L$_{FUV}$ & IRX & M$_{dust}$ & $\beta_{calz94}$ & SFR$_{10Myrs}$ & f$_{burst}$ &  age$_{burst}$ & $\tau_{main}$ & age$_{main}$ & $\delta$ & M$_{star}$ & A$_{FUV}$ \\
            $\tau_{500Myrs}$\_dl2014 & 0.98 & 0.89 & 0.91 & 0.98 & 0.89 & 0.96 &  &  &  & 0.81 &  & 0.95 & 0.92 \\
            $\tau_{500Myrs}$\_casey2012\_OT & 0.98 & 0.84 & 0.87 & 0.98 & 0.84 & 0.96 &  &  &  & 0.80 &  & 0.93 & 0.85 \\
            $\tau_{20000Myrs}$\_dl2014 & 0.97 & 0.92 & 0.89 & 0.97 & 0.87 & 0.95  &  &  &  & 0.71 &  & 0.94 & 0.89 \\            
            $\tau_{20000Myrs}$\_casey2012\_OT & 0.98 & 0.90 & 0.89 & 0.98 & 0.87 & 0.97 &  &  &  & 0.74 &  & 0.95 & 0.89 \\
            $\tau_{500Myrs}$\_dl2014\_burst & 0.96 & 0.90 & 0.88 & 0.96 & 0.91 & 0.93 & 0.47 & 0.05 &  & 0.46 &  & 0.96 & 0.93 \\
            $\tau_{500Myrs}$\_casey2012\_OT\_burst & 0.96 & 0.94 & 0.90 & 0.96 & 0.89 & 0.91 & 0.27 & 0.06 &  & 0.42 &  & 0.94 & 0.92 \\
            $\tau_{multi}$\_dl2014 & 0.98 & 0.56 & 0.87 & 0.98 & 0.73 & 0.96 &  &  & 0.20 & 0.65 &  & 0.13 & 0.90 \\
            $\tau_{multi}$\_casey2012\_OT & 0.98 & 0.64 & 0.88 & 0.98 & 0.80 & 0.97 &  &  & 0.18 & 0.62 &  & 0.19 & 0.91 \\
            $\tau_{500Myrs}$\_dl2014\_delta & 0.99 & 0.93 & 0.90 & 0.99 & 0.85 & 0.97 &  &  &  & 0.78 & 0.37 & 0.94 & 0.89 \\
            $\tau_{500Myrs}$\_casey2012\_OT\_delta & 0.98 & 0.92 & 0.79 & 0.98 & 0.56 & 0.97 &  &  &  & 0.79 & 0.35 & 0.92 & 0.62 \\
            \noalign{\smallskip}
            \hline
            \noalign{\smallskip}
        \end{tabular}}
      \caption[]{Summary of the power of two for correlation coefficients, $r^2$, from the mock analysis on the main physical parameters performed by CIGALE. The delayed  SFH  with $\tau_{main}$ = 500 Myrs is shown with ”tau500” in the  legend. If a final  burst was added, ”burst” is used in the legend. A constant SFH with $\tau_{main}$ = 20 Gyrs is shown with "tau20000" in the legend, and multi-$\tau_{main}$ is shown as ”multitau” in the legend.}
         \label{Tab.Mocks}
   \end{table*}

   \begin{table*}
   \begin{center}
 
         $\begin{array}{c|c|c}
            \hline
            \noalign{\smallskip}
            Parameter & DL2014 & PL+OT\_MBB \\
            \noalign{\smallskip}
            \hline
            \noalign{\smallskip}
            M_{star}\ [M_{\odot}] & (1.18 \pm 0.19) \times 10^{10} & (1.18 \pm 0.18) \times 10^{10} \\
            SFR\ [M_{\odot}\ yr^{-1}] & 47.6 \pm 5.1 & 46.8 \pm 5.0 \\
            {M_{dust}\ [M_{\odot}]}^* & (5.64 \pm 0.69) \times 10^{7} & (2.07 \pm 0.25) \times 10^{7} \\
            L_{dust}\ [L_{\odot}] & (3.38 \pm 0.41) \times 10^{11} & (3.32 \pm 0.40) \times 10^{11} \\
            L_{FUV}\ [L_{\odot}] & (1.07 \pm 0.06) \times 10^{11} & (1.06 \pm 0.06) \times 10^{11} \\
            sSFR\ [yr^{-1}] & (1.16 \pm 0.34) \times 10^{-8} & (1.15 \pm 0.34) \times 10^{-8} \\
            {sM_{dust}}^* & 0.010 \pm 0.003 & 0.004 \pm 0.001 \\
            IRX & 0.31 \pm 0.14 & 0.31 \pm 0.13 \\
            A_{FUV} & 1.14 \pm 0.15 & 1.14 \pm 0.15 \\
            age_{main}\ [Myr] & 440 \pm 98 & 444 \pm 98 \\
             \noalign{\smallskip}
            \hline
         \end{array}$

   \end{center}
     \caption[]{Mean value and standard deviations of the physical parameters derived from the fit analysis of the individual objects in our sample. The Bayesian outputs are listed here. $^*$The values listed in the table are those computed with $\kappa_0$ = 0.637 m$^2 kg^{-1}$, corresponding to \cite{Draine2014}. For $\kappa_0$ = 0.45 m$^2 kg^{-1}$ [$\kappa_0$ = 0.72 m$^2 kg^{-1}$, respectively], that is before [after, respectively] the reverse shock, we found the following: $M_{dust}$ = (2.91 $\pm$ 0.35) $\times 10^{7}$\ M$_{\odot}$\  [$M_{dust}\ $ = (1.82 $\pm$ 0.22) $\times 10^{7}\ M_{\odot}$, respectively] and $sM_{dust}$ = 0.005 $\pm$ 0.002 [$sM_{dust}$ = 0.003 $\pm$ 0.001, respectively].}
      \label{Tab.Results_CIGALE.Ph3}
   \end{table*}




\section{Analysis of the results}

From the previous analysis, we derived a set of physical parameters for each of the galaxies in our sample. These parameters allowed us to define and build diagnostic diagrams that permitted us to characterize these galaxies and, more specifically, their SFH and their dust properties. We analyzed the locations of our sample in the IRX versus $\beta_{UV}$ diagram , the A$_{FUV}$ versus M$_{star}$ diagram, and in the specific dust mass versus specific star formation rate: sM$_{dust}$ versus sSFR = M$_{dust}$ / M$_{star}$ versus SFR / M$_{star}$ (dust formation rate diagram or DFRD).

\subsection{The SFR versus M$_{star}$ diagram}
\begin{figure*}
   \centering
       \includegraphics[angle=0,width=15cm]{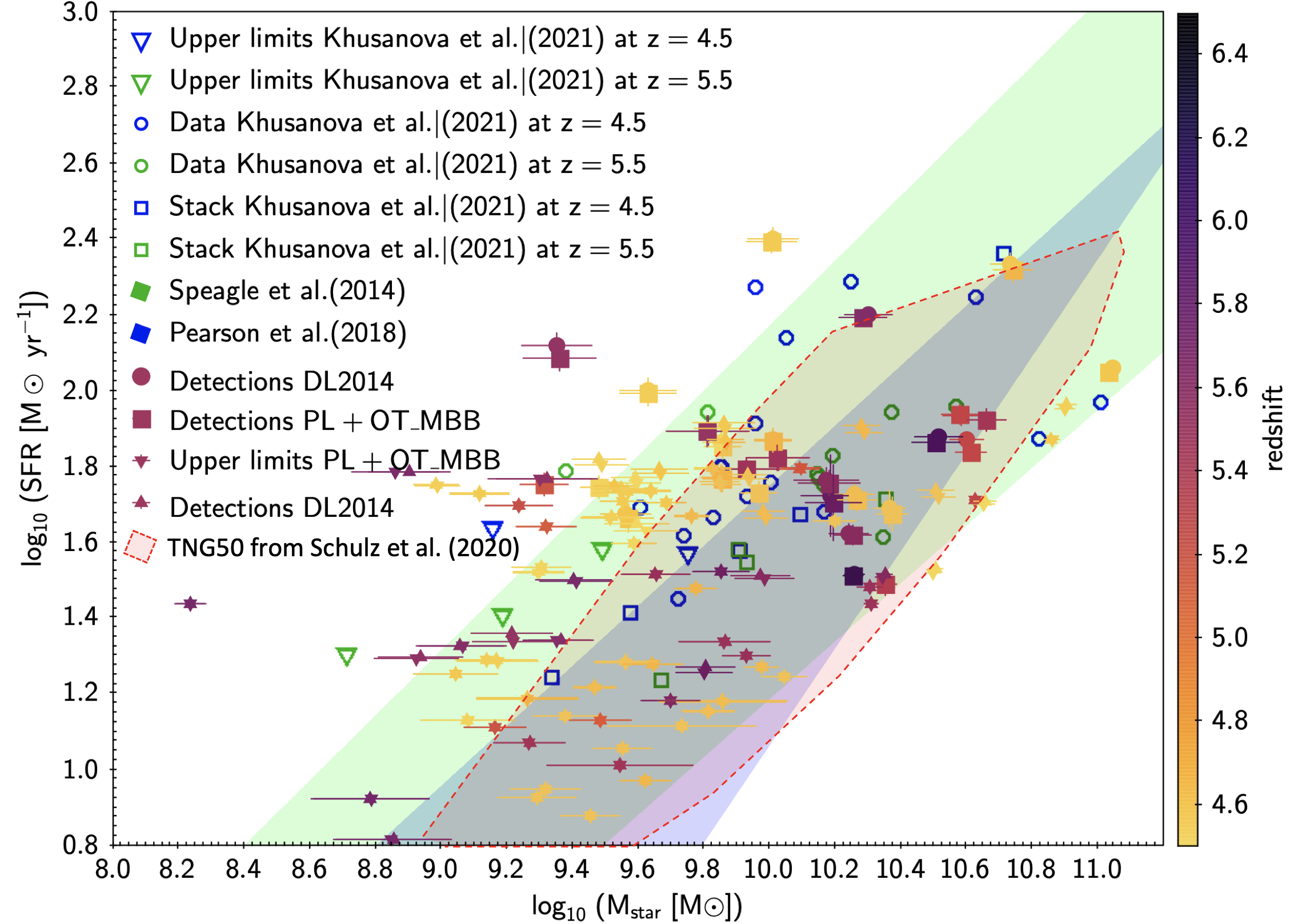}
   \caption{In the $\log_{10}$ (SFR) vs. $\log_{10}$ (M$_{star}$) diagram, the main part of the sample is found within the limits for the fits of the main sequence at z = 5.0 by \citet{Speagle2014} (green shading), by \citet{Pearson2018} at z $\sim$ 5.2 (purple shading), and by \citet{Faisst2020a}, who found that the galaxies are in agreement with \citet{Speagle2014}. Is is important to note, however, that some of our objects are at redshifts larger than the 4.5 - 5.6 ALPINE sample (see color code of the markers). For \cite{Speagle2014}, we used the “mixed” (preferred fit)'' function (as defined by \cite{Speagle2014}). The results from the two fits with DL2014 and PL+OT\_MBB are presented. The two types of dust emission do not significantly modify the location of the points in the diagram. Detections are shown as dots and boxes and upper limits are shown as downward- and upward-pointing triangles. The uncertainty range for upper limits extends to the bottom of the plot. In addition to having mainly upper limits, at $\log_{10}$ (M$_{star})<9.5$, the sample is very likely incomplete which means that it is difficult to estimate a trend from these data over the entire mass range. We also added the objects and stacks from \citet{Khusanova2021} with open markers. Finally, the selection of TNG50 galaxies used in \cite{Schulz2020} is also provided (red-shaded area).}
   \label{Fig.SFR_Mstar}
   \end{figure*}

Fig.~\ref{Fig.SFR_Mstar} presents the SFR versus M$_{star}$ diagram\footnote{We only present the results using the reference SFH derived in the present section, i.e., delayed SFR(t) = $t/\tau_{main}^2$ exp(t/$\tau$) with $\tau_{main}$ = 500Myrs, but with the following two options for dust emission: \citet{Draine2014} and a power law in a mid-IR and optically thin blackbody.}. The points corresponding to this work are found in the expected range when compared to Pearson et al. (2018) who also used CIGALE: 
$$\log_{10}(SFR) = (1.00 \pm 0.22)\ (log_{10}(M_{star}) - 10.6) + (1.92 \pm 0.21)$$ 
and compared to \cite{Speagle2014} with their ''mixed” (preferred fit)'': 
$$\log_{10}(SFR) = [(0.73 \pm 0.02) - (0.027 \pm 0.006) \times t[Gyr] ] \log_{10}(M_{star})$$
$$- [(5.42 \pm 0.22) + (0.42 \pm 0.07)\times t[Gyr] )]$$ 
function evaluated at z = 5.0 (i.e., t$_{universe}$ = 1.186 Gyr), which was converted to a Chabrier IMF by subtracting 0.03 to $\log_{10}$(SFR) and to $\log_{10}$(M$_{star}$). The first result is that regardless of the dust emission used (DL2014 or PL+OT\_MBB), we found about the same main sequence. Our data are in good agreement with \citet{Faisst2020a} and \cite{Khusanova2021} and also they generally follow the relations derived by \citet{Speagle2014} and \cite{Pearson2018}. Most of the detections are found at a relatively large stellar mass ($\log_{10}(M_{star}) \sim 10$). We confirm the evolution of the main sequence to z = 4.5 - 5.5 with our sample of galaxies (mainly objects not detected in dust continuum), which extend to $\log_{10}(M_{star}) \sim 8.5$.

\subsection{The A$_{FUV}$ versus M$_{star}$ diagram}

The relation between the dust attenuation (A$_{FUV}$ or its proxy, IRX = L$_{dust}$ / L$_{FUV}$) and the stellar mass (M$_{star}$) is another way to estimate the dust attenuation in galaxies without far-IR data. This relation between the stellar mass and dust attenuation has been the focus of numerous studies \citep[as early as][and references therein]{Xu2007, Buat2009}. However, even if this relation could be very useful in addition to the IRX versus $\beta_{FUV}$ one (f$_{\lambda}$ $\propto$ $\lambda~^{\beta_{FUV}}$), the link between the far-UV (FUV) dust attenuation and the stellar mass (A$_{FUV}$ versus M$_{star}$) is not well established at all redshifts.

\cite{Bouwens2016} define what a consensus relation could be: $\log_{10}$(IRX) = $\log_{10}$(M$_{star}$/M$_\odot$) - 9.17 assuming the dust temperature evolves with the redshift. This relation is linear in the plane $\log_{10}$(IRX) versus $\log_{10}$ (M$_{star}$/M$_\odot$), and in the range 9.0 $<$ $\log_{10}$ (M$_{star}$/M$_\odot$) $<$ 11.0. In a recent paper, \cite{Carvajal2020} used the stacking method for 1582 UV LBGs with photometric redshifts in the range z $\sim$ 2 - 8 to reach down to $\log_{10}$ (M$_{star}$/M$_\odot$) = 6.0. However, the constraints from this stacking are only upper limits, which are less useful than detections (see their Fig.~16). \cite{Fudamoto2020} made use of the ALPINE data and show that the $\log_{10}$(IRX) versus M$_{star}$ relation derived from their observations is inconsistent with the previously determined relations at z $\leq$ 4. They found a fast decrease in IRX at z $\sim$ 4 in massive galaxies which suggests an evolution of the average amount of dust attenuation in star forming galaxies. \cite{Bernhard2014} assume an evolving normalization of the $\log_{10}$(IRX) versus $\log_{10}(M_{star})$ relation in the low redshift universe at z $<$ 1. We need an opposite evolution (lower in IRX) in the high redshift universe at z $>$ 4 as in \cite{Bogdanoska2020}.

\begin{figure*}[h!]
   \centering
       \includegraphics[angle=0,width=9cm, height=5cm]{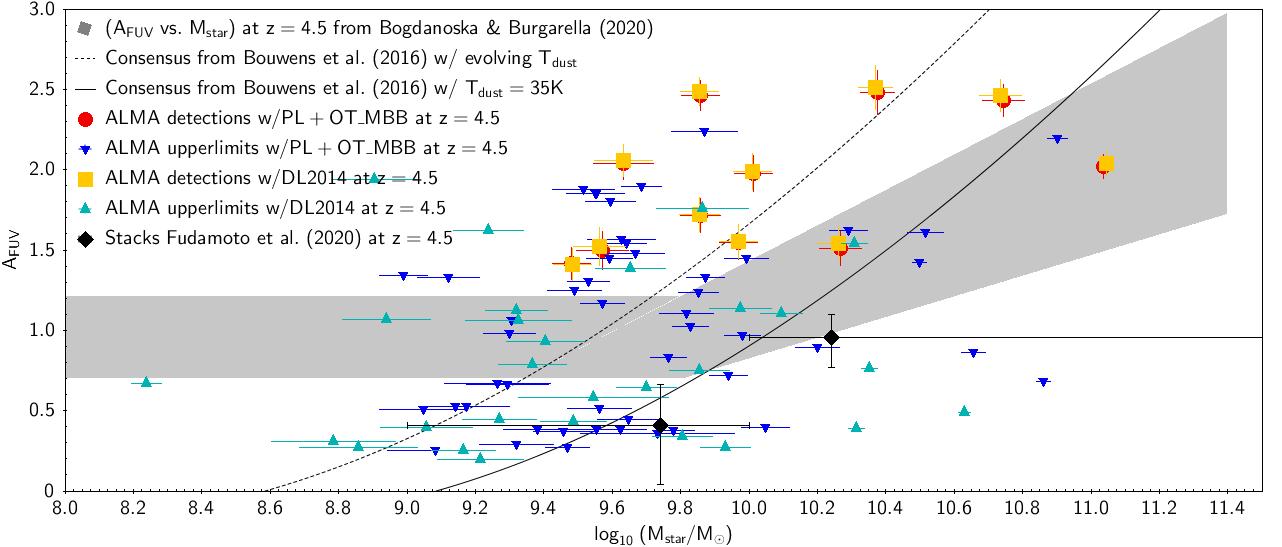}
       \includegraphics[angle=0,width=9cm, height=5cm]{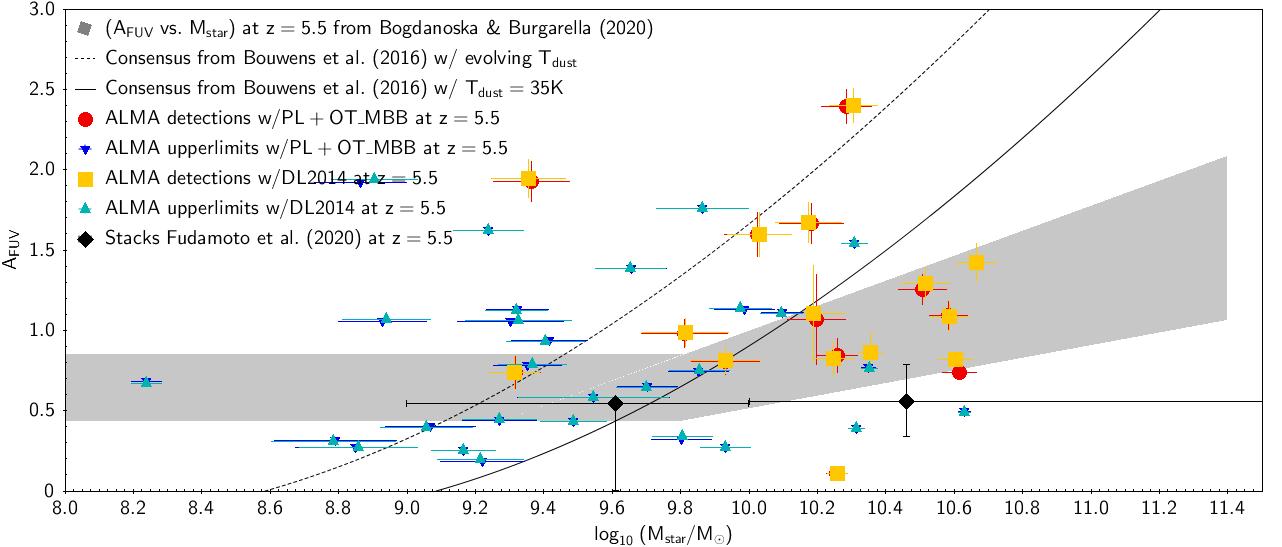}
   \caption{A$_{FUV}$ - M$_{star}$ diagram at z $\sim$ 4.5 (left) and z $\sim$ 5.5 (right). The gray areas correspond to the expected relation at z = 4.5 and 5.5 from \cite{Bogdanoska2020}. This relation was formed by a broken line, which is flat at $\log_{10}$ M$_{star}$ $\leq$ 9.8 and rises at $\log_{10}$ (M$_{star}$) $>$ 9.8. The ALPINE data are very dispersed at z $\sim$ 4.5, while this flatness is supported by the data at z $\sim$ 5.5. The conversion from IRX to A$_{FUV}$ is from Burgarella et al. (2005): A$_{FUV}$ = -0.028 [$\log_{10}$ (IRX)]$^3$ + 0.392 [$\log_{10}$ (IRX)]$^2$ + 1.094 [$\log_{10}$ (IRX)] + 0.5.}
      \label{Fig.AFUV_Mstar}
\end{figure*}

Fig.~\ref{Fig.AFUV_Mstar} shows that the two M$_{star}$ stacks from \cite{Fudamoto2020} at z $\sim$ 4.5 are marginally (accounting for the uncertainties) in agreement with \cite{Bogdanoska2020}. They are in better agreement at z $\sim$ 5.5. However, the low stellar mass range is crucial, especially in the early universe where galaxies are expected to have lower stellar masses because they were still building their stars at that time. From our analysis, we confirm the effect found by \cite{Fudamoto2020}: our galaxies have lower A$_{FUV}$ or IRX values at a fixed mass, compared to previously studied IRX–M$_{star}$ relations at z $<$ 4, with quite a large scatter. 

In order to constrain the amount of dust attenuation of these low-mass galaxies, it is possible to adopt a global approach and compare what the hypothesis on the A$_{FUV}$ versus $\log_{10}$ (M$_{star}$) implies for the redshift evolution of the average dust attenuation in the universe as presented in \cite{Cucciati2012, Burgarella2013, Madau2014}, for example. Using this approach, \cite{Bogdanoska2020} conclude that it is not possible to extend a linear relation to the lowest mass range without strongly underpredicting the average dust attenuation in the universe at all redshifts.  We need a flattening of the relation at low stellar mass ($\log_{10}$(M$_{star}$/M$_\odot$) $<$ 9.0). This means that the apparent dust attenuation of the low-mass galaxies is significantly higher than 0. This very interesting and unexplained point is also suggested by other observational and theoretical works \citep[e.g., ][]{Salim2016, Takeuchi2010, Cousin2019, Ma2016}. They propose modeling the A$_{FUV}$ versus M$_{star}$ relation with the following broken law:
\begin{equation}
 A_{FUV} =
    \begin{cases}
      \text{a $\times$ 1.1} & \text{at $\log_{10}$M$_{star} <$ 9.8}\\
      \text{a $\times$ ($\log_{10}$M$_{star} - 8.7$)} & \text{at $\log_{10}$M$_{star} \geq 9.8 $}\\
    \end{cases}       
\end{equation}
where a = (z + $\gamma$) $\times$ $\alpha\ ^{\beta - (z+\gamma)}$ with z being the redshift and the constants $\alpha$ = 1.84 $\pm$ 0.11, $\beta$ = 1.84 $\pm$ 0.12, and $\gamma$ = 0.14 $\pm$ 0.04. Although there is some dispersion and the completeness is very likely small at $\log_{10}$M$_{star} < 10.0$, Fig.~\ref{Fig.AFUV_Mstar} suggests that the relation from \cite{Bogdanoska2020} at z = 5.0 is in broad agreement with the data. However, the large number of upper limits at low mass could suggest that the level of the flat relation might be overestimated with respect to the present data. Alternate ways to measure A$_{FUV}$ for low-mass galaxies should be explored, maybe via the Balmer decrements with JWST or with radio data using the radio-to-IR ratio $q_{IR}$ that allows one to constrain the far-IR emission of star-forming galaxies from radio data \citep[e.g., ][]{Helou1985, Delvecchio2021}.


\subsection{The IRX - $\beta_{FUV}$ diagram}

\begin{figure*}[h!]
   \centering
      \includegraphics[angle=0,width=18cm]{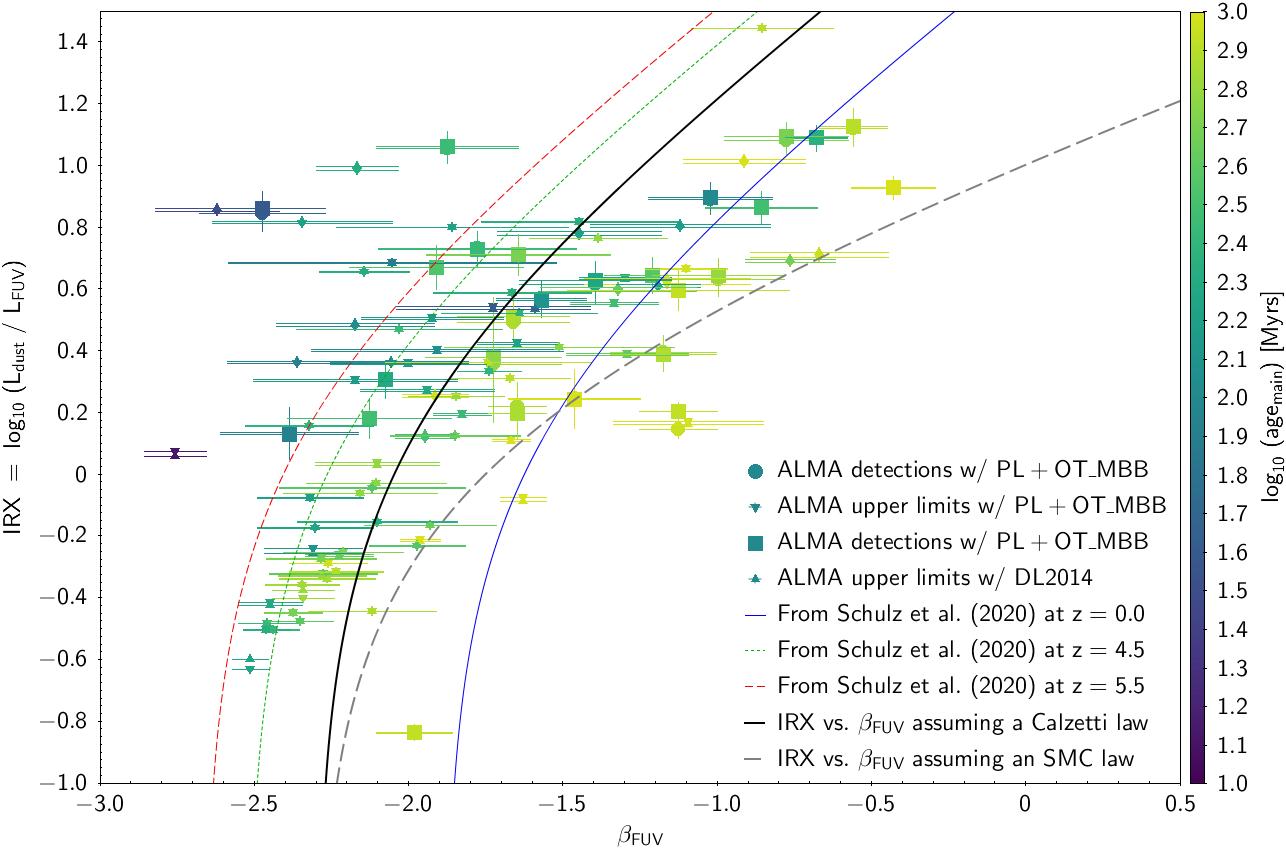}
   \caption{IRX - $\beta_{CFUV}$ diagram. IRX values were estimated with CIGALE and $\beta_{FUV}$ were fitted on the data directly. They are not model-dependant. The black continuous line corresponds to the original Calzetti law, under the assumption that the underlying dust curve follows the Calzetti et al. (2000) attenuation. The dashed line to the predicted law assumes the SMC extinction law (e.g., Gordon et al. 2003). Both are from \cite{McLure2018}. The color of the symbols are related to the axis to the right of the figure, the age of the stellar population. Both results with dust emission, DL2014 and PL+OT\_MBB, are presented with different symbols with ALMA upper limits and ALMA detections. In the figures, we also plotted the laws from \cite{Schulz2020} at z = 0.0 (blue continuous line), 4.5 (green dotted line), and 5.5 (red dashed line). A comparison with the IRX - $\beta_{FUV}$ plot in \cite{Fudamoto2020} with the same sample suggests that our (especially ALMA-detected) galaxies extend less to very blue $\beta_{FUV}$. This is mainly true for the subsample of galaxies not detected in continuum with ALMA. This is probably due to two effects: first, in \cite{Fudamoto2020}, 3$\sigma$ upper limits are plotted. Another possible effect could be because we used a unique IR composite template in the SED fitting, which reduces the uncertainties on L$_{dust}$.}
   \label{Fig.IRX-beta}
   \end{figure*}

To estimate the UV slope $\beta_{FUV}$, we used the definition given by \cite{Calzetti1994} in the range 125-260~nm within ten selected windows (see their Tab. 2) designed to remove all absorption features and the 217.5~nm dust bump ($\beta_{Calzetti-1994}$) from the fitting procedure. The IRX - $\beta_{FUV}$ is a classical tool to estimate the dust attenuation of galaxies. Fig.~\ref{Fig.IRX-beta} shows the location of the galaxies in the IRX versus $\beta_{FUV}$ diagram. We also show the classical positions assuming a Calzetti law and a small Magellanic cloud (SMC) law. Our galaxies appear to be systematically shifted to the left of the diagram, that is to bluer values of $\beta_{FUV}$ compared to the locus estimated by \cite{Overzier2010} for galaxies at z = 0. This might be related to the evolution of the stellar population at z $\sim$ 4.5 - 5.0 as they were younger and bluer (and/or an intrinsic evolution of other parameters similar to the IMF). Age effects were already found at low redshift \cite[e.g.,][]{Kong2004, Boquien2009}. 

The IllustrisTNG Project \citep[TNG hereafter:][]{Nelson2018, Pillepich2018} is a suite of cosmological simulations for the formation of galaxies. TNG50 does not model dust directly, but \cite{Schulz2020} assume that the diffuse dust content of a galaxy is traced by the gas-phase metal distribution assigned to this galaxy. The dust density distribution is derived from the TNG50 gas density distribution with assumptions on the dust-to-metal ratio, the gas metallicity, the gas temperature, and the instantaneous SFR. Finally, SKIRT \citep{Baes2011} is used to model the emission of the galaxies. Their fiducial model is a multicomponent dust mix, which models a composition of graphite, silicate, and polycyclic aromatic hydrocarbon (PAH) grains, with various grain size bins for each grain type which reproduce the properties of Milky Way (MW), large Magellanic cloud (LMC), and SMC type dust. In SKIRT, the dust of the molecular birth clouds mentioned is treated separately from the diffuse ISM dust.

\cite{Schulz2020} used the output of the TNG50 simulation and suggest a redshift-dependent systematic shift toward lower $\beta_{FUV}$ with increasing redshift modeled by adding a component $\beta_z (z) = 0.142z\ -\ 0.081$ to the $\beta_{FUV}$ value at z=0. The trend was calibrated up to z = 4. If we extrapolate their trend out to the redshift of our galaxies, the corresponding locus in Fig.~\ref{Fig.IRX-beta} would appear too blue when compared to most of our galaxies. In order to extend the redshift range, we computed $\beta_z = \beta - \beta_{Overzier\ et\ al.\ 2010}$ because \cite{Overzier2010} is the reference at z = 0 used by \cite{Schulz2020}. 

\begin{figure*}
   \centering
       \includegraphics[angle=0,width=20cm]{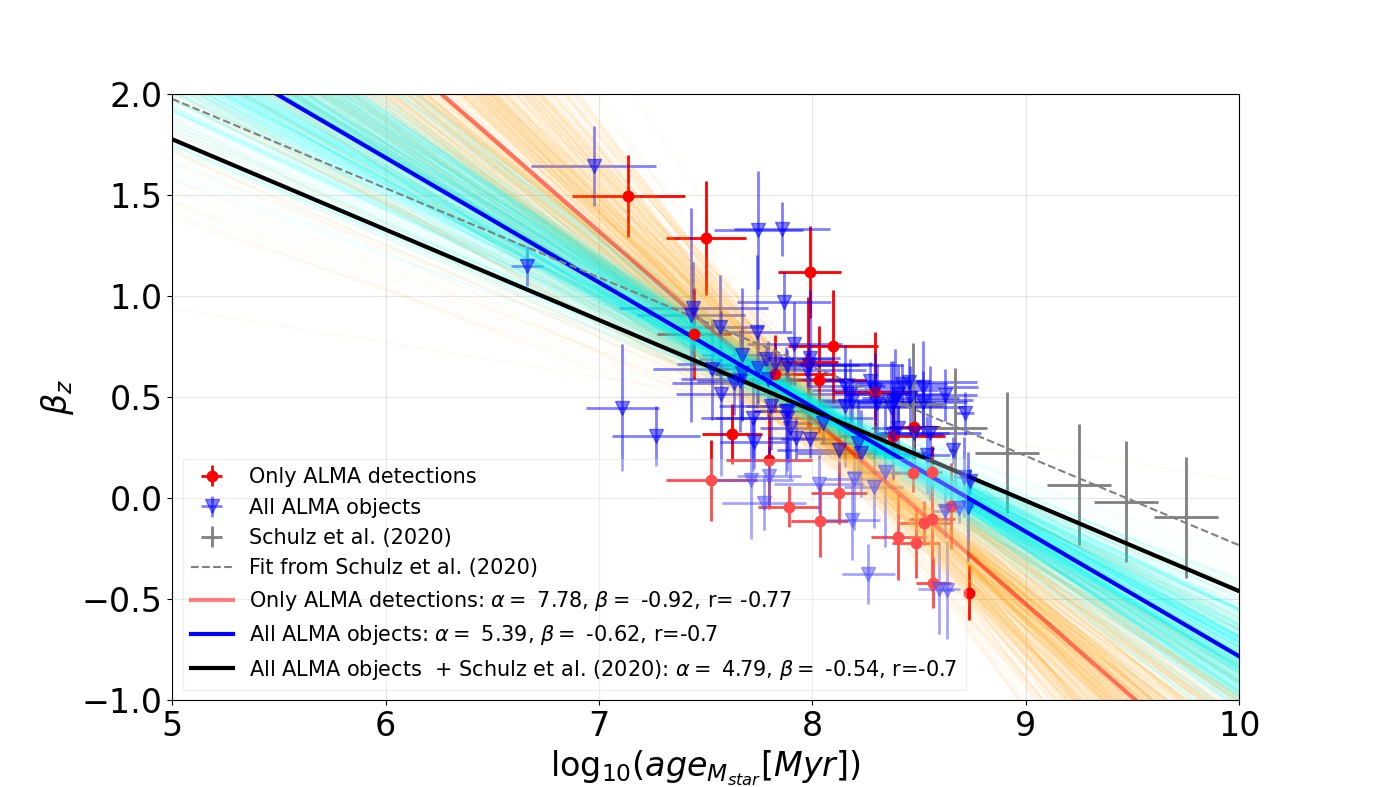}
   \caption{Linear relation between the shift of the IRX-$\beta_{FUV}$ relation at z=0 (namely that of \cite{Overzier2010}) vs. the mass-weighted age in years derived from the analysis of the galaxies studied here. Red symbols correspond to ALMA-detected objects, while blue objects correspond to upper limits from ALMA. We note, however, that these upper limits apply to ALMA, but not to the parameters presented in this figure: $age_{M_{star}}$ and $\beta_z$.}
   \label{Fig.beta_z_logage_M_star}
   \end{figure*}

An evolution in redshift of this effect is observationally confirmed. 
One of the main parameters that can be at the origin of the shift is likely to be the evolution of the stellar population, as mentioned above. So, even though using the redshift as the independent variable might be a simple solution, a more realistic and physical approach should be related to the stellar populations themselves. \cite{Schulz2020} present, in their Figure~7 (middle), the variations of the intrinsic UV slopes $\beta_0$ of the galaxy stellar population (i.e., before the attenuation by dust) against their mass weighted mean stellar population ages. This figure shows that $\beta_0$ correlates with the stellar population age: the younger the stellar population, the lower  $\beta_0$ is. From Figure 7 (middle) from \cite{Schulz2020}, we measured the mean and uncertainties over each axis of the intrinsic UV–slopes $\beta_0$ and of the mass-weighted mean stellar population ages (age\_M$_{star}$) for each redshift range. CIGALE provides mass-weighted mean stellar population ages for each of the galaxies and for our two best models defined above (i.e., delayed SFH with $\tau=500$Myrs plus DL2014 and PL+OT\_MBB for the IR emission). These values are plotted in Fig.~\ref{Fig.beta_z_logage_M_star}. The average in the four bins of  age\_M$_{star}$ from \cite{Schulz2020} and our data points seem to follow a linear relation. Although the points from \cite{Schulz2020} are not fully compatible with ours, there is a general trend. Statistical tests with the LINMIX library, which uses a hierarchical Bayesian approach for the linear regression with an error in both X and Y \citep{Kelly2007}, suggest that the trend is highly significant (see correlation coefficient in Fig.~\ref{Fig.beta_z_logage_M_star}), given the number of points used. The equations in Fig.~\ref{Fig.beta_z_logage_M_star} allow one to quantify this systematic shift along $\beta_{FUV}$ at large redshift.

\subsection{The dust formation rate diagram (DFRD)}

\cite{Burgarella2020} and \cite{Nanni2020} built a diagram to follow the evolution of the dust mass called the dust formation rate diagram (DFRD) which shows the specific dust mass (M$_{dust}$ / M$_{star}$ = sM$_{dust}$) versus the specific star formation rate (SFR / M$_{star}$ = sSFR). The specific dust mass was already identified by \cite{Calura2017} as a quantity that represents a true measure of how much dust per unit stellar mass survives the various destruction processes in galaxies. However, this is also a quantity that allows one to quantify the various dust formation processes. The interpretation of this DFRD is that galaxies would follow a dust cycle where they start to build their dust grains at high sSFR leading to a fast rise in sM$_{dust}$. After this phase, the galaxies would reach a maximum in sM$_{dust}$ before losing their dust grains and going down in the DFRD. At the end of this dust cycle, \cite{Burgarella2020} and \cite{Nanni2020} found that about 80\% of the mass fraction of the total baryons had been removed by the outflow and the rest had mainly been destroyed by supernovae (SNe). However, we still have difficulty understanding the evolutionary status of galaxies with very high sM$_{dust} > 0.01$. \cite{Calura2017} propose that spiral galaxies are characterized by a nearly constant sM$_{dust}$ as a function of the stellar mass and cosmic time, whereas proto-spheroids present an early steep increase of the sM$_{dust}$, which stops at a maximal value and decreases in the latest stages.

We built this same diagram from the large galaxy sample studied in this paper and for our two main models (delayed SFH with $\tau=500$ Myrs plus DL2014 and PL+OT\_MBB for the IR emission). When using the module DL2014 in CIGALE, the dust mass is provided with the models. We note that the models of DL2014 assume an optically thin dust. The CIGALE module casey2012\_OT, that is PL+OT\_MBB, does not give M$_{dust}$, and we need to compute it. 

The IR SED
$$ S_\nu \propto \nu^{\beta_{RJ}} B_{\nu}(T_{dust})$$

was computed with a modified blackbody
$$ B_{\nu}(T_{dust}) = \frac{2h}{c^2}\frac{\nu^3}{e\ ^{\frac{h\nu}{kT_{dust}}}-1}$$

and the dust mass was derived with the following formula:
$$ M_{dust}\ =\ \frac{L_{\nu}} {4\pi \kappa_{\nu} B_{\nu}(T_{dust})}$$

where $\nu$ = c / $\lambda_{200\ \mu m}$, h is the Planck constant, and c is the speed of light. We note that the emission of the CMB was neglected given the dust temperatures T$_{} \gtrsim 40$K found in this paper \citep{daCunha2013}.

In the models for DL2014, the dust composition \citep{Weingartner2001} corresponds to 30\% of graphite and 70\% of silicate (MgFeSiO4). We first adopted the constant given in \cite{Draine2003} at $\lambda=200~\mu$m: $\kappa_0$ = 0.637\ m$^2$ kg$^{-1}$ because we also assumed $\lambda_0$ = 200 $\mu$m, $\kappa_{\nu}=\kappa_0\ (\lambda_0\ / \lambda)\ ^\beta\ = \kappa_0$. 
When comparing M$_{dust}$(DL2014) to M$_{dust}$(PL+OT\_MBB), we found the latter to be systematically smaller by a factor of 0.37 $\pm$ 0.01. This was already noticed in several works \citep{Magdis2013, Santini2014, Bianchi2019, Magnelli2012}. The origin of this shift might be found in the single-temperature models that are unable to account for the wide range in the temperature of dust grains that are exposed to different intensities of the interstellar radiation field \citep[see][ for a  very pedagogical description]{Liang2019}.

\begin{figure*}
   \centering

   \includegraphics[angle=0,width=16.5cm]{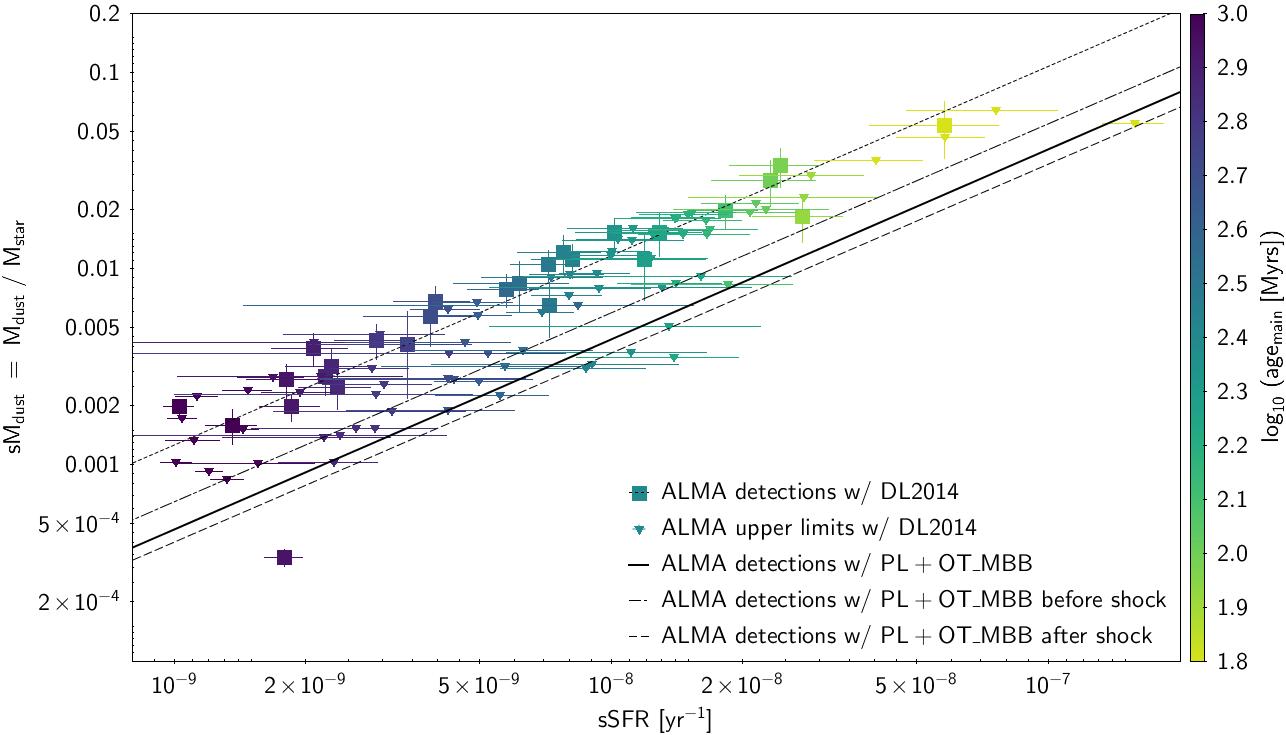}
   \includegraphics[angle=0,width=16.5cm]{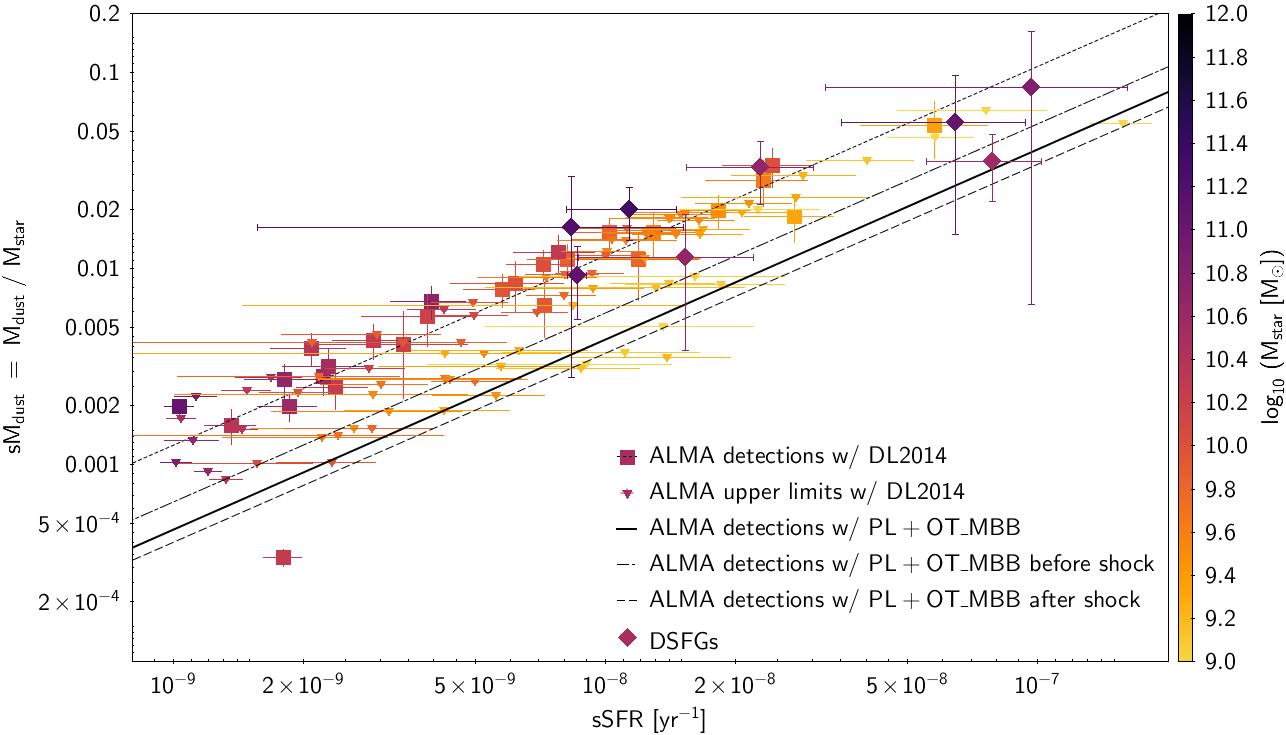}

   \caption{Comparison of observed DFRDs with models. Top: DFRDs color-coded with $\log_{10}$age$_{main}$. The individual symbols were computed assuming a PL+OT\_MBB emission. We clearly observe an age sequence from right to left. The lines show the trend assuming different dust emissions and different values for $\kappa_0$. The top one assumes DL2014 dust emission. The second one from the top was derived when using an optically thin modified blackbody with $\kappa_0$ = 0.637 m$^2$ kg$^{-1}$ (PL+OT\_MBB). We note that the factor 0.37 needed to match DL2014 to PL+OT\_MBB emission was not applied. The third one from the top assumes optically thin modified blackbodies with $\kappa_0$ corresponding to SNe dust mass absorption coefficients. It corresponds to the situation before the SNe reverse shock destruction, and the bottom line represents after the reverse shock destruction. Both from \cite{Hirashita2017}. Bottom: Same models as in as (a). The same objects plotted in (a) are color-coded in M$_{star}$. A sample of DSFGs (Tab.~\ref{Tab.DSFGs}), also color-coded in M$_{star}$, was added to the plot with the code as follows: the upper triangle shows the maximum value, and the lower triangle shows the minimum value. These high redshift DSFGs are found on the same sequence as the other objects. However, these DSFGs have stellar masses larger than the underlying galaxy population that we study. It is important to note that the physical parameters (especially M$_{star}$ because the IR emission is dominant) of these high redshift DSFGs have very large uncertainties and their position location in the diagram can almost cover the entire plot. Better estimates of these parameters coming from JWST would help.}
  \label{Fig.DFRDs}
\end{figure*}

Fig.~\ref{Fig.DFRDs} presents the DFRD with data points color-coded with the parameters age$_{main}$. Only individual results obtained using DL2014 are shown. However, the trends are also presented for other dust emissions (PL+OT\_MBB with various assumptions on $\kappa_0$, as explained below). The lines are based on a fit with only the ALMA-detected objects. We see an age sequence from right to left, that is with decreasing sSFR. 

We note that $\kappa_0$ = 0.637\ m$^2$ kg$^{-1}$ corresponds to the dust opacity in the models of DL2014. In the early universe, most if not all of the dust could only be produced by SNe \citep[e.g.,][]{Burgarella2020, Nanni2020}. \cite{Hirashita2017} propose values for $\kappa_0$ for dust condensed in SNe before reverse shock destruction, $\kappa_{158\ \mu m}$ = 0.557 m$^2$ kg$^{-1}$, and for dust ejected from SNe after reverse shock destruction, $\kappa_{158\ \mu m}$ = 0.894 m$^2$ kg$^{-1}$. The dust composition and grain size distribution assumed in \cite{Hirashita2017} are from \cite{Nozawa2003}. After correction to get these dust mass absorption coefficients at $\lambda = 200$ $\mu$m using $\kappa_{\nu}\ =\ \kappa_0\ (\lambda_0\ / \lambda)\ ^{\beta_{RJ}}$ with $\beta_{RJ}$ from the SED fitting (Tab.~\ref{Tab.fit_IR_template}), we obtained $\kappa_0$ = 0.45 and 0.72 m$^2$ kg$^{-1}$. This means that, when adopting the SNe value for $\kappa_0$, the dust masses would be roughly of the same order as PL+OT\_MBB with $\kappa_0$ = 0.637\ m$^2$ kg$^{-1}$. This is certainly a crucial point: given the age of these galaxies, it is very likely that the dust grains have been produced by SNe. So we could wonder how to reconcile this assumption with the fact that the present models are far from being able to reproduce diagnostic diagrams such as Fig.~\ref{Fig.DFRDs}. This point is also debated from a laboratory standpoint. For instance, \cite{Fanciullo2020, Ysard2019} suggest that current dust masses are overestimated by up to a factor of 10 - 20 or 2 - 5, depending on the assumptions on grain structure (porous or compact, respectively). Laboratory measurements of dust analogs show that FIR opacities, that is to say mass absorption coefficients, are usually higher than the values used in models and that they depend on several parameters, including temperature, composition, shape, and morphology, for instance. This would mean that dust mass estimates may be overestimated. The properties (e.g., dust composition and grain size) are still unknown for galaxies at a very large redshift. This is important when deriving the dust mass \citep[e.g.,][]{Ysard2019, Hirashita2017, Inoue2020}.  

\begin{figure*}
   \centering
     \includegraphics[angle=0,width=15cm]{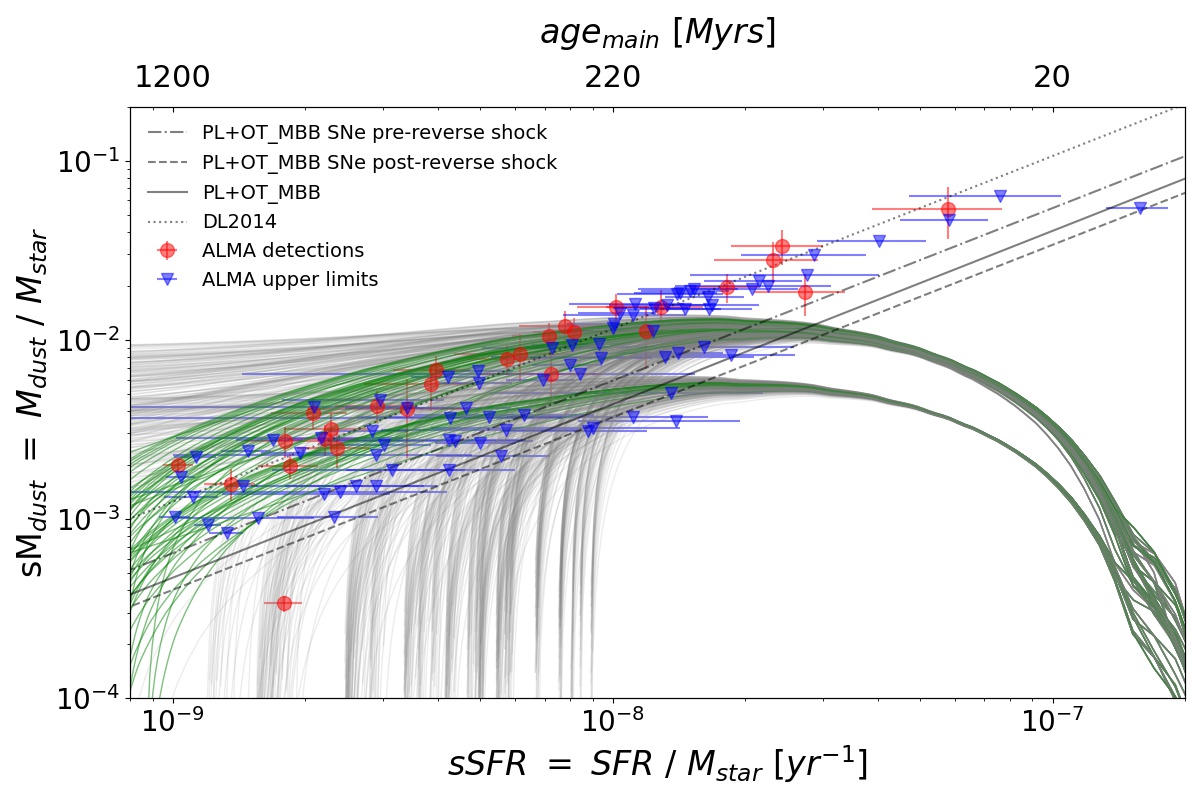}
     \label{Fig.DFRD_wmodels.a}
   \caption{DFRD for models with dust masses computed by CIGALE with DL2014 dust models. Only models with a delayed star formation history are plotted. A delayed star formation history only was consistently used in the SED fitting with CIGALE to estimate sM$_{dust}$ and sSFR. The models shown do not have any grain growth in the ISM. We also show the other models in light gray to show which parameter space the models cover. On this
figure, we superimposed the three lines that show the respective linear fits as in Fig.~\ref{Fig.DFRDs}. The models shown are from \cite{Burgarella2020} and \cite{Nanni2020}. This figure shows that most of the objects (ALMA detections and ALMA upper limits) are approximately located at the same positions as the models. Only HZ9 is found at larger sSFR and sM$_{dust}$. If confirmed at this location in the DFRD, these objects would be very difficult to explain with these models.}
   \label{Fig.DFRD_wmodels}
   \end{figure*}

One interesting point, which is beyond of the scope of this work, is related to the nature of objects that are usually considered to belong to another class: the high redshift dusty star-forming galaxies (DSFG) such as ADFS-27 \citep{Riechers2021} or HFLS3 \citep{Riechers2013}. The very massive and very dusty galaxies have been found in the high redshift universe thanks to wide-field far-IR and submm observations with, for example, the South Pole Telescope or the European Space Agency's Herschel Space Observatory. High redshift DSFGs are at the high mass end, which suggests that they are not similar to the early phases of the SFGs analyzed here. In an attempt to check their location in the DFRD, we added the DFSGs from Tab.~\ref{Tab.DSFGs} to Fig.~\ref{Fig.DFRD_wmodels}. All of them fall on the same sequence identified for our studied sample; however, they do not show the same stellar mass as our sample.  We can wonder whether the stellar mass of these high redshift DSFGs are badly estimated or whether they follow a different path. It should be noticed that CIGALE with similar assumptions on the SFH and on the dust emission has been used for some of these objects, such as ADFS-27. 

\begin{table*}
  \centering
  \small
\begin{tabular}{|l|c|c|r|r|r|r|r|r|r|r|r|l|r|r|r|r|}
\hline
  \multicolumn{1}{|c|}{id} &
  \multicolumn{1}{c|}{z} &
  \multicolumn{1}{c|}{$\mu$} &
  \multicolumn{1}{c|}{M$_{dust}$} &
  \multicolumn{1}{c|}{M$_{dust}$\_err} &
  \multicolumn{1}{c|}{M$_{star}$} &
  \multicolumn{1}{c|}{M$_{star}$\_err} &
  \multicolumn{1}{c|}{M$_{gas}$} &
  \multicolumn{1}{c|}{M$_{gas}$\_err} &
  \multicolumn{1}{c|}{SFR} &
  \multicolumn{1}{c|}{SFR\_err} &
  \multicolumn{1}{c|}{f$_{gas}$$^*$} &
  \multicolumn{1}{c|}{ref} \\
\hline
  ADFS-27      & 5.7 & 1.0 & 4.2E9 & 0.4E9 & 2.1E11 & 0.6E11$^{\ \ }$ & 2.1E11 & 0.2E11 & 2380 &  230 & 0.50 & $^1$ \\
  ADFS-27 N    & 5.7 & 1.0 & 2.6E9 & 0.4E9 & 1.6E11 & 1.3E11$^{\ \ }$ & 1.2E11 & 0.1E11 & 1330 &  130 & 0.43 & $^1$ \\
  ADFS-27 S    & 5.7 & 1.0 & 1.5E9 & 0.2E9 & 4.6E10 & 1.5E10$^{\ \ }$ & 9.4E10 & 0.8E10 & 1050 &  110 & 0.66 & $^1$ \\
  HFLS-3       & 6.3 & 2.2 & 1.3E9 & 0.3E9 & 3.7E10 & 1.1E10$^{\ \#}$ & 1.0E11 & 0.1E11 & 2900 &  180 & 0.73 & $^2$,$^3$ \\
  SPT 0311-58$^{\ \$}$  & 6.9 & 2.0 & 6.1E9 & 3.5E9 & 1.1E11 & 0.5E11$^{\ \ }$ & 6.4E11 & 3.7E11 & 7082 & 3967 & 0.85 & $^4$,$^5$ \\
  SPT 0311-58W & 6.9 & 2.2 & 5.5E9 & 3.5E9 & 6.6E10 & 4.4E10$^{\ \ }$ & 5.9E11 & 3.7E11 & 6380 & 3960 & 0.90 & $^4$,$^5$ \\
  SPT 0311-58E & 6.9 & 1.3 & 5.2E8 & 2.6E8 & 4.6E10 & 2.0E10$^{\ \ }$ & 5.2E10 & 2.6E10 &  702 &  228 & 0.53 & $^4$,$^5$ \\
  GN10         & 5.3 & 1.0 & 1.1E9 & 4.4E8 & 1.2E11 & 0.6E10$^{\ \ }$ & 7.1E10 & 0.9E10 & 1030 &  190 & 0.37 & $^6$ \\

\hline
    \end{tabular}
    \caption[]{Physical parameters for the sample of high redshift dusty star-forming galaxies. We note that the parameters were not corrected for the gravitational magnification. $^1$: \citet{Riechers2021}, $^2$: \citet{Riechers2013}, $^3$ Cooray2014, $^4$ Strandet2017; $^5$ Marrone2018, $^6$: \citet{Riechers2020}. * f$_{gas}$ = M$_{gas}$/(M$_{gas}$+M$_{star}$) were computed from the listed M$_{gas}$ and M$_{star}$ values. $^{\#}$ No error quoted, we assume 30\%. $^{\$}$ The parameters for SPT031158 are the sum of the two components (W \& E). }
  \label{Tab.DSFGs}
\end{table*}

In Fig.~\ref{Fig.DFRD_wmodels}, we compare the entire sample with the models built in \citet{Burgarella2020} and in \citet{Nanni2020}. The models are fully explained in \cite{Nanni2020}. In brief, we performed the calculations for the metal evolution using the One-zone Model
for the Evolution of GAlaxies (OMEGA) code (\citealt{Cote2017}). We assumed the metal yields for type II SNe from \citet{Kobayashi2006} computed up to 40 M$_{\odot}$, and from the FRUITY database for low-mass stars with M > 1.3 M$_{\odot}$ evolving through the thermally pulsing AGB phase and developing stellar winds (\citealt{Cristallo2011}, \citealt{Piersanti2013}, \citealt{Cristallo2015}). The yields for population III stars are  from \citet{Heger2010} and are limited to the mass range 10 < M/M$_{\odot}$ < 30 in OMEGA. Dust removal from the galaxy through galactic outflow follows a rate proportional to the SFR through the "mass-loading factor": ML $\times$ SFR. This assumes that the galactic outflow is generated by the feedback of stars on the gas in the ISM (e.g., \citealt{Murray2005}). Two kinds of IMFs were tested: top-heavy IMFs and a Chabrier IMF. We assume that a fraction of silicates (olivine and pyroxene), iron, and carbon grains ejected in the ISM are condensed. The best models selected by \cite{Burgarella2020} generally agree with the Alpine sample. Only one object (HZ9 at sSFR $\sim$ 5 - 6 $\times$ 10$^{-8}$ yr$^{-1}$) lies at high sSFR. This is consistent with a fast evolution of these young objects, regardless of their sM$_{dust}$. With the present data, we cannot rule out that the trend could keep rising to the right of the diagram.  Galaxies have to start building dust grains early in their lifetime, at $\log_{10 }$ sSFR $\geq$ 10$^{-8}$ - 10$^{-7}$ yr$^{-1}$, to the right of the DFRD. It therefore seems important to measure dust and stellar masses for galaxies at high sSFR and young ages to locate the locus of this rising sM$_{dust}$ and thus bring some observational data to constrain the models. 

\section{Conclusions}

This paper studies a sample of star-forming galaxies observed with ALMA in the redshift range 4.5 $\lesssim$ z $\lesssim$ 6.2. Some of them, detected in the continuum with ALMA, were used to build a composite IR SED covering a large wavelength range from about 15 $\mu$m to 218 $\mu$m. Using CIGALE, we modeled the IR composite template and derived a set of physical properties. Building on the assumption that this IR composite template is valid for an ensemble of galaxies selected in the same way, such as ALPINE's SFGs and LBGs, we use it to fit the SEDs of the ALPINE sample plus the sample of LBGs from \cite{Burgarella2020} and \cite{Nanni2020} from the far-UV to the submm ranges.

The following results were found:
   \begin{itemize}
      \item We built a unique z$>$4 IR SED. It is compatible with the ALPINE sample completed with galaxies from \cite{Burgarella2020} plus the z $\sim$4.5 and z $\sim$5.5 stacks from \cite{Bethermin2020}. We provide a table to the readers with the raw composite IR SED and the modeled SEDs.
      \item Except for an SFH based on a delayed with a final burst history, all other SFHs (constant, delayed with $\tau_{main}$ = 500 Myrs, and delayed with several $\tau_{main}$) are in agreement with the data. For simplicity, we selected the delayed SFH with $\tau_{main}$ = 500 Myrs.
      \item We checked the position of the sample in the SFR versus M$_{star}$ diagram. The objects in the sample follow the same trend as those previously derived in papers at the same redshift.
      \item When comparing the position of the sample with the evolution in redshift of the IRX versus M$_{star}$ relation found by \cite{Bogdanoska2020} at z $\sim$ 4.5 and at z $\sim$ 5.5, we found a reasonable agreement. This agreement especially holds at z $\sim$ 5.5. However, the absolute level of this flat relation must be reevaluated using detections of low-mass galaxies.
      \item The sample studied in this paper was placed in the IRX versus $\beta_{FUV}$ diagram. An evolution in age is observed with younger galaxies having bluer $\beta_{FUV}$.
      \item Moreover, our sample was found to be shifted to bluer $\beta_{FUV}$ with respect to a z = 0 reference. We modeled this evolution with the mass-weighted age and provide an equation that allows one to predict the position of the sequence as a function of the redshift.
      \item We plotted our galaxies in the DFRD (M$_{dust}$/M$_{star}$ versis SFR/M$_{star}$, DFRD). The objects form a sequence that can be described by evolution in various time-dependent parameters, and more specifically the age of the stellar populations and the stellar mass. This suggests that the sequence is due to the evolution of the galaxies over cosmic time.
      \item The models built by \cite{Burgarella2020, Nanni2020} are generally in agreement with the data, although the sample is much larger than in these previous papers. 
      \item High redshift, high mass DSFGs over-plotted in the DFRD are found on the same sequence followed by our sample. However, the stellar mass estimated for these high redshift DSFGs are larger than the one found at the same locus for our sample. Uncertainties on the stellar mass of these objects are notoriously high because the stellar emission is faint, but the disagreement seems large. Better estimates of their stellar masses should be available if they are observed with JWST.
   \end{itemize}

\begin{acknowledgements} This program receives funding from the CNRS national program Cosmology and Galaxies. D.R. acknowledges support from the National Science Foundation under grant No. AST-1910107. D.R. also acknowledges support from the Alexander von Humboldt Foundation through a Humboldt Research Fellowship for Experienced Researchers. M.T. acknowledges the support from grant PRIN MIUR 2017 20173ML3WW 001. A.N. acknowledges support from the Narodowe Centrum Nauki (UMO-2018/30/E/ST9/00082 and UMO-2020/38/E/ST9/00077). G.C.J. acknowledges ERC Advanced Grant 695671 ``QUENCH'' and support by the Science and Technology Facilities Council (STFC). Y.F. further acknowledges support from NAOJ ALMA Scientific Research Grant number 2020-16B. M.R. acknowledges support from the Narodowe Centrum Nauki (UMO-2020/38/E/ST9/00077). Médéric Boquien gratefully acknowledges support by the ANID BASAL project FB210003 and from the FONDECYT regular grant 1211000.  E.I.\ acknowledges partial support from FONDECYT through grant N$^\circ$\,1171710

\end{acknowledgements}

%
%

\bibliographystyle{aa}
\bibliography{ALPINE}

\appendix

\section{CIGALE parameters for the initial fits}\label{Appendix.A}
\subsection{Star formation histories (SFHs)}
       \begin{itemize}
           \item [$\ast$] A delayed SFH: SFR(t) = $t / \tau_{main}^2\ \exp(-t/\tau_{main})$ with $\tau_{main}$ in the range 25 to 10000 Myrs and main ages ($t$ or $age_{main}$) in the range 2 - 1200 Myrs is the first option. This type of SFH allows for SFRs to increase or decrease, depending on the age of the stellar population.
           \item [$\ast$] The second option is a delayed SFH and a final young burst with the burst age ($age_{burst}$) are in the range 2 to 50 Myrs, but main ages are in the range 100 to 1200 Myrs, and the fraction of burst ($f_{burst}$) is as follows: 0.0, 0.001, 0.01, 0.10, 0.50 and $\tau_{burst}$ = 20000Myrs, that is to say it is similar to a constant SFH given the burst ages.
           \item [$\ast$] Fixed parameters: because it was not possible to estimate $\tau_{main}$ with confidence, we also sequentially fixed and tried $\tau_{main}$ = 25, 250, and 2500 Myrs both for the delayed and burst runs before comparing the results to the above runs.
       \end{itemize}
\subsection{Dust characteristics}
       \begin{itemize}
           \item [$\ast$] The models from \cite{Draine2014}: because no data are available in the rest-frame near-IR and mid-IR, we could not constrain the mass fraction of PAH (q$_{PAH}$). However, it is generally accepted \citep[e.g., ][]{Douglas2010,  Castellano2014, Yuan2019, Bellstedt2021} that the metallicity of these galaxies are likely subsolar. \cite{Ciesla2014} show a relation between q$_{PAH}$ and the metallicity. Without any other constraint, we arbitrarily chose a low value for the mass fraction of PAH: q$_{PAH}$=0.47. The impact of this parameter on the dust mass or luminosity is very low. We used the full range of allowed parameters for the minimum of the distribution of the starlight intensity relative to the local interstellar radiation field, U$_{min}$, for $\alpha$, the power law slope dU/dM = U$^\alpha$, as well as for $\gamma$, the fraction of the dust heated by starlight above the lower cutoff U$_{min}$. For the same reason as for $\tau_{main}$, in the SFH parameters, we also sequentially tried fixed values for $\alpha$ = 1.0, 2.0, and 3.0. 
           \item [$\ast$] Modified blackbody plus a power law similar to \cite{Casey2012}: Our initial priors are the dust temperature in the range 30K $\leq$ T$_{dust}$ $\leq$ 85K and the emissivity in the RJ part of the SED 0.5 $\leq$ $\beta_{RJ}$ $\leq$ 2.0. We kept the MIR power slope at the default $\alpha_{MIR}$ = 2.0 because no data are available to constrain $\alpha_{MIR}$. These values are consistent with \cite{Faisst2020b}, that is 40 $<$ T$_{dust}$[K] $<$ 60 (called  T$_{SED}$, i.e., SED dust temperature in their paper), with a median at 48 K and emissivity indices ($\beta_d$ in their paper) between 1.6 and 2.4 for all galaxies, and a median of 2.0. As before, because it could not be safely estimated with the data available, we tried several fixed values for $\beta_{RJ}$= 1.0, 1.5, and 2.0.
           \item [$\ast$] Other parameters: In addition to these parameters, we selected a Chabrier IMF and a wide range of dust attenuation. We selected CIGALE's dust attenuation law from the Calzetti and Leitherer module. It is based on the \cite{Calzetti2000} starburst attenuation curve, extended with the \cite{Leitherer2002} curve between the Lyman break and 120 nm. A nebular contribution was also added \citep[see][]{Boquien2019} where $D_\lambda$ is the Drude profile, and the last term renormalizes the curve so that E(B-V) remains equal to the input E(B-V) when $\delta$ is not 0:
   
                            $$k_\lambda=k_\lambda^{starburst}\times(\lambda/\lambda_V)^\delta\times D_\lambda)\frac{E(B-V)_{\delta=0}}{E(B-V)_\delta}$$.
       \end{itemize}
   
   In Boquien et al. (submitted), a variation of the dust attenuation law is studied. However, changes in the shape of the dust attenuation law do not impact the IRX versus A$_{FUV}$ relation for star forming galaxies because it is almost completely independent of the extinction mechanisms (i.e., dust and star geometry, attenuation law \citep[e.g., ][and references therein]{Witt2000, Cortese2008}.
   
   Finally, from these initial fits, we examined and discarded the fits for which the reduced $\chi_\nu^2 \geq 5.0 $ for each
of our detected objects. These bad fits only represent a few of them with respect to all of the attempts. For each object in the sample, we computed the mean normalization factor from the CIGALE Bayesian luminosity at 200 $\mu$m: L$_{200\mu m}$. After applying this normalization, all the observed SEDs started to share the same flux density at $\lambda$ = 200 $\mu m$: f$_{\nu}(200\ \mu$m) = 1.0 (see Fig.\ref{Fig.Flowchart}) and we could then proceed to the next phase to try and combine all the observed SEDs into a single one.

\section{CIGALE parameters for the final fit}
\begin{table*}[htp]
\begin{center}
 \resizebox{0.8\linewidth}{\height}{
\begin{tabular}{|>{\centering}p{6.0cm}|>{\centering\arraybackslash}p{3.0cm}|>{\centering\arraybackslash}p{2.5cm}|>{\centering\arraybackslash}p{2.5cm}|}
 
  \hline\hline
  {\bf Parameters} & {\bf Symbol} & {\bf Range (Ph.~2, DL2014)} & {\bf Range (Ph.~2, PL+OT\_MBB)} \\
  \hline
 Target  sample    &                      &  Individual Hi-z LBGs & Individual Hi-z LBGs  \\ 
  \hline\hline
    \multicolumn{4}{c}{}\\
    \multicolumn{4}{c}{\bf Delayed SFH and recent burst}\\
  \hline
 e-folding time scale of the delayed SFH & $\tau_{main}$ [Myr] & 500 & 500 \\ 
  \hline
 Age of the main population & Age$_{main}$[Myr]  & 101 log values in [2 - 1200] & 101 log values in [2 - 1200]  \\ 
  \hline
 Burst & f$_{burst}$  &  No burst  &  No burst  \\ 
  \hline
    \multicolumn{4}{c}{}\\
    \multicolumn{4}{c}{\bf SSP}\\
  \hline
  SSP &   & BC03 & BC03 \\ 
  \hline
  Initial mass function &  IMF & Chabrier & Chabrier \\ 
  \hline
  Metallicity     & Z &  0.004 &  0.004 \\ \hline
    \multicolumn{4}{c}{}\\
    \multicolumn{4}{c}{\bf Nebular emission}\\
  \hline
  Ionization parameter &  logU    & -2.5, -2.0, -1.5 & -2.5, -2.0, -1.5 \\
  Line width [km/s]    &     ---    &  100 &  100 \\
  \hline
    \multicolumn{4}{c}{}\\
    \multicolumn{4}{c}{\bf Dust attenuation law}\\
  \hline
  Color excess for both the old and young stellar populations &  E\_BV\_lines & 101 log values in [0.01, 1.0] & 101 log values in [0.01, 1.0] \\ 
  \hline
  Bump amplitude &  uv\_bump\_amplitude &  0.0 & 0.0 \\ 
  \hline
  Power law slope & power law\_slope & 0.0 & 0.0 \\ 
  \hline
    \multicolumn{4}{c}{}\\
    \multicolumn{4}{c}{\bf Dust emission (DL2014)}\\
  \hline
  Mass fraction of PAH & $q_{PAH}$ &  0.47 & ---  \\ 
  \hline
  Minimum radiation field &  U$_{min}$ & 17.0 & ---  \\ 
  \hline
  Power law slope dU/dM $\approx$ U$^\alpha$ &---  $\alpha$ & 2.4 &  \\ 
  \hline
   Dust fraction in PDRs  & $\gamma$ & 0.54 &  ---  \\ 
  \hline
    \hline
    \multicolumn{4}{c}{}\\
    \multicolumn{4}{c}{\bf Dust emission (PL + OT\_MBB)}\\
  \hline
  Dust temperature & $T_{dust}$ & ---  & 54.1 \\ 
  \hline
  Emissivity &  $\beta_{RJ}$ & ---  & 0.87 \\ 
  \hline
   Slope of the MIR power law  & $\alpha_{MIR}$ & ---  & 2.0\\ 
  \hline
    \multicolumn{4}{c}{}\\
    \multicolumn{4}{c}{\bf No AGN emission}\\
\hline\hline
\end{tabular}}
  \caption{CIGALE modules and input parameters used for all the fits. BC03 means \cite{Bruzual2003}, and the Chabrier IMF refers to \cite{Chabrier2003}.}
  \label{Tab.CIGALE.Ph2}
\end{center}
\end{table*}

\section{Results for the mock analysis performed with CIGALE}
   \begin{figure*}
   \centering

   \begin{subfigure}{\textwidth}
       \includegraphics[angle=0,width=18.5cm]{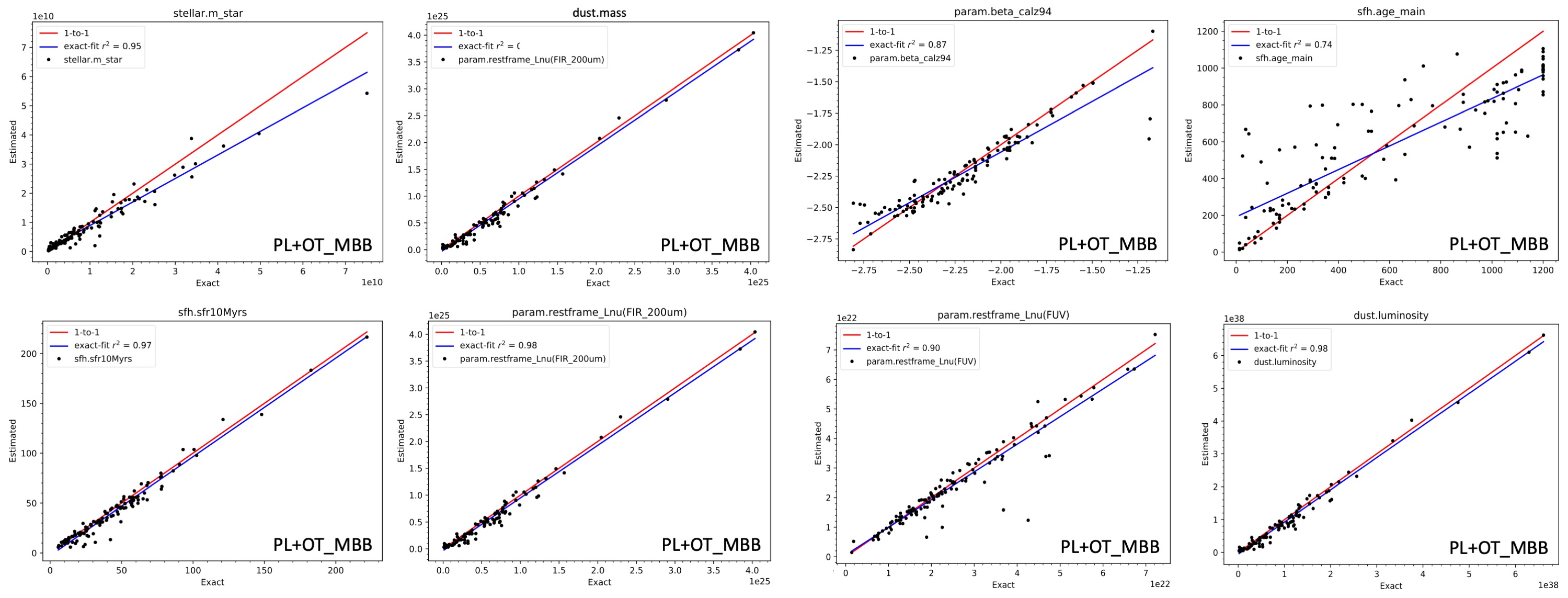}
       \caption{Mock analysis using the power law and OT\_MBB in the mid-IR as in Casey et al. (2012) for the IR template.} \label{Fig.mocks_OT_MBB}
   \end{subfigure}
   \begin{subfigure}{\textwidth}
   \includegraphics[angle=0,width=18.5cm]{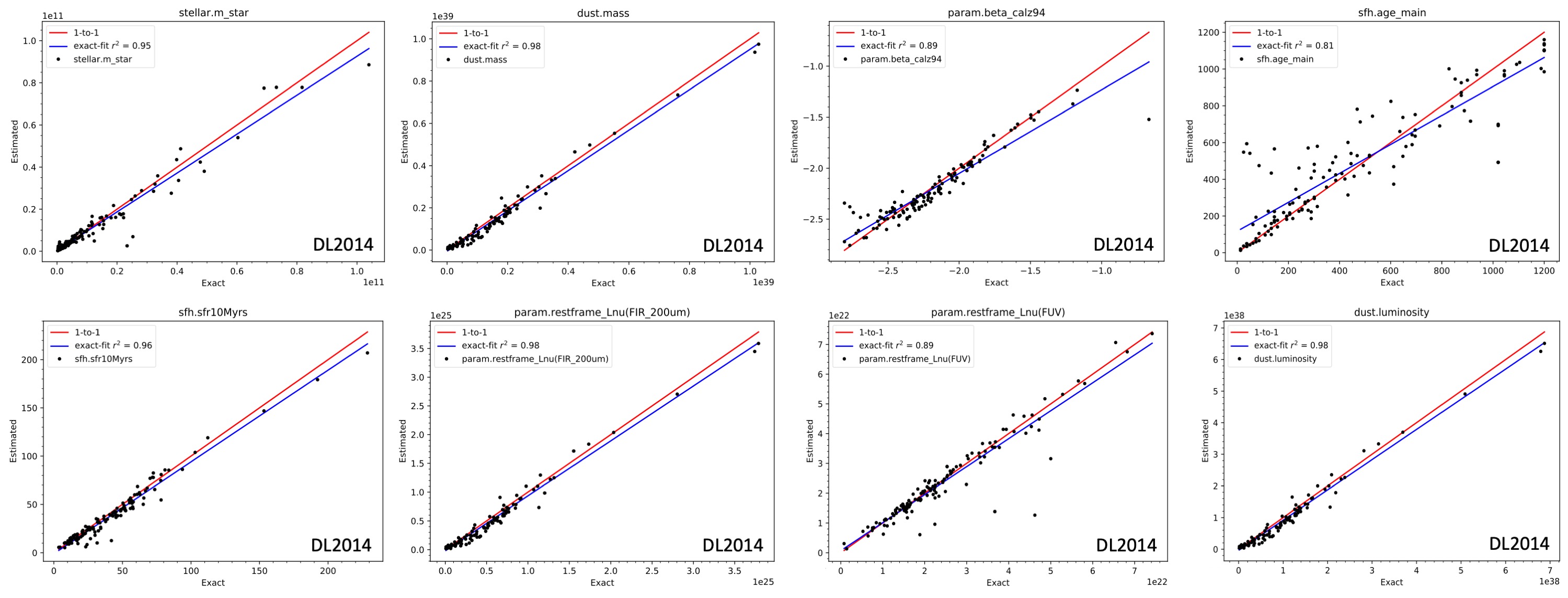}
       \caption{Mock analysis using the DL2014 model for the IR template.} 
       \label{Fig.mocks_dl2014}
   \end{subfigure}
   \caption{Results from the mock analysis performed with CIGALE where we compared input parameters (i.e., parameters from the best fit for each object) to the same parameters estimated by CIGALE with the very same priors used for the true analysis \citep[see ][]{Boquien2019}. We note that some parameters were estimated from another one via a linear relation, e.g., M$_{dust}$ from L$_{200 \mu m}$, which explains why the figures are identical.}
   \label{Fig.Mocks}
   \end{figure*}



\onecolumn

\section{The observed IR composite template}

\sisetup{table-column-width=12ex,    
         round-mode=places,
         round-precision=3,
         tight-spacing,
         table-format = 5.3e-2,
         table-number-alignment = center
         }

\begin{longtable}[c]{|S|S|S|S|}
 
 \caption{IR template for each of the models, normalized to 200 $\mu$m. Only the first rows are given here.}
 \label{Tab.SED_IR.composite.models}\\

 \hline
 \multicolumn{4}{| c |}{Beginning of table}\\
 \hline
    \multicolumn{1}{|c|}{$\lambda[\mu m]$} &
    \multicolumn{1}{c|}{F$_{\nu}$(DL2014)} &
    \multicolumn{1}{c|}{F$_{\nu}$(PL+OTMBB)} &
    \multicolumn{1}{c|}{F$_{\nu}$(PL+GMBB)} \\
 \hline
 \endfirsthead

 \hline
 \multicolumn{4}{| c |}{Continuation of Table~\ref{Tab.SED_IR.composite.models}}\\
 \hline
    \multicolumn{1}{|c|}{$\lambda[\mu m]$} &
    \multicolumn{1}{c|}{F$_{\nu}$(DL2014)} &
    \multicolumn{1}{c|}{F$_{\nu}$(PL+OTMBB)} &
    \multicolumn{1}{c|}{F$_{\nu}$(PL+GMBB)} \\
 \hline
 \endhead

 \hline
 \multicolumn{4}{| c |}{End of Table}\\
 \hline\hline
 \endlastfoot

\hline
  20.0 & 0.18107749474964585 & 0.033371661115243596 & 0.22818696797630056\\
  20.14 & 0.18551481313133614 & 0.03558839180505059 & 0.23284902077654243\\
  20.32 & 0.1916912479528318 & 0.038600899856686154 & 0.23899800048124503\\
  20.51 & 0.19834306021901293 & 0.0419773937900743 & 0.2456753421353187\\
  20.7 & 0.20531825078431543 & 0.0455757038111942 & 0.25255705038272314\\
  20.867 & 0.21165231238228988 & 0.048922292048504665 & 0.2587737366722555\\
  20.89 & 0.21253443514863027 & 0.04939717591545251 & 0.2596425809753937\\
  21.09 & 0.21999878126005848 & 0.05366062340145448 & 0.26731997222221277\\
\end{longtable}



\section{Physical parameters for each galaxy of the studied sample}

\sisetup{table-column-width=12ex,    
         round-mode=places,
         round-precision=3,
         tight-spacing,
         table-format = 5.3e-2,
         table-number-alignment = center
         }
  \begin{table}[h!]
  \centering
\begin{adjustbox}{max width=\textwidth}

\begin{tabular}{|l|S|S|S|S|S|S|S|S|S|S|S|S|S|S|S|S|S|S|S|S|S|S|S|S|S|S|S|S|S|l|S|S|S|S|S|S|S|S|S|S|S|S|S|S|S|S|S|S|S|S|S|S|}

\hline
  \multicolumn{1}{|c|}{id} &
  \multicolumn{1}{c|}{redshift} &
  \multicolumn{1}{c|}{$\beta_{FUV}$} &
  \multicolumn{1}{c|}{$\beta_{FUV}$\_err} &
  \multicolumn{1}{c|}{sSFR} &
  \multicolumn{1}{c|}{sSFR\_err} &
  \multicolumn{1}{c|}{sM$_{dust\_dl2014}$} &
  \multicolumn{1}{c|}{sM$_{dust\_dl2014}$\_err} &
  \multicolumn{1}{c|}{sM$_{dust\_pre-shock}$} &
  \multicolumn{1}{c|}{sM$_{dust\_pre-shock}$\_err} &
  \multicolumn{1}{c|}{sM$_{dust\_post-shock}$} &
  \multicolumn{1}{c|}{sM$_{dust\_post-shock}$\_err} &
  \multicolumn{1}{c|}{M$_{dust\_dl2014}$} &
  \multicolumn{1}{c|}{M$_{dust\_dl2014}$\_err} &
  \multicolumn{1}{c|}{M$_{dust\_pre-shock}$} &
  \multicolumn{1}{c|}{M$_{dust\_pre-shock}$\_err} &
  \multicolumn{1}{c|}{M$_{dust\_post-shock}$} &
  \multicolumn{1}{c|}{M$_{dust\_post-shock}$\_err} &
  \multicolumn{1}{c|}{A$_{FUV}$} &
  \multicolumn{1}{c|}{A$_{FUV}$\_err} &
  \multicolumn{1}{c|}{IRX} &
  \multicolumn{1}{c|}{IRX\_err} &
  \multicolumn{1}{c|}{age$_{main}$} &
  \multicolumn{1}{c|}{age$_{main}$\_err} &
  \multicolumn{1}{c|}{L$_{dust}$} &
  \multicolumn{1}{c|}{L$_{dust}$\_err} &
  \multicolumn{1}{c|}{L$_{FUV}$} &
  \multicolumn{1}{c|}{L$_{FUV}$\_err} &
  \multicolumn{1}{c|}{SFR } &
  \multicolumn{1}{c|}{SFR\_err} &
  \multicolumn{1}{c|}{M$_{star}$} &
  \multicolumn{1}{c|}{M$_{star}$\_err} \\
     &   &   &   & [yr$^{-1}$] & [yr$^{-1}$] &   &   &   &   &   &   & [Myrs] & [Myrs] & [L$_\odot$] &  & [M$_\odot$] & [M$_\odot$] & [L$_\odot$] & [L$_\odot$] & [M$_\odot$ yr$^{-1}$] & [M$_\odot$ yr$^{-1}$] & [M$_\odot$ yr$^{-1}$] & [M$_\odot$ yr$^{-1}$] \\
\hline
  CANDELS\_GOODSS\_12 & 4.431000232696533 & -2.144344430007398 & 0.1460967634012939 & 1.0069640335966491E-8 & 3.062395929299864E-9 & 0.004715939430851186 & 0.0014341273933664788 & 0.006620897832979376 & 0.002013429368672506 & 0.0041251007751336825 & 0.0012544520786919304 & 1.9960376874831546E7 & 2060778.414705241 & 2.8023179248544946E7 & 2893210.046531481 & 1.7459631813690756E7 & 1802592.8365973544 & 1.5665890629827568 & 0.09418145081174756 & 0.6520640523959013 & 0.05005426413638634 & 225.65478612399394 & 73.74800516536834 & 3.193129856753339E11 & 3.297188265305342E10 & 7.080738181452225E10 & 3.540369090726112E9 & 42.62010126445132 & 4.402695914241293 & 4.2325346131998286E9 & 1.2106744448488061E9\\
   CANDELS\_GOODSS\_14 & 5.552700042724609 & -1.9627663677483977 & 0.06621954990145637 & 1.3261226412638939E-9 & 1.154215751295264E-10 & 3.160246818134731E-4 & 7.18346280159452E-5 & 4.4367981431202714E-4 & 1.008515355079121E-4 & 2.764313831899319E-4 & 6.283479337573494E-5 & 6498864.565893009 & 1414214.7236769209 & 9124018.43833292 & 1985473.167468151 & 5684651.308894224 & 1237034.1770467116 & 0.3912359041277432 & 0.07600263936782348 & -0.21201247306620347 & 0.10750851771179427 & 1028.2639631966056 & 20.583247293827203 & 1.0384652472279529E11 & 2.2636185114695347E10 & 1.6498025003058682E11 & 8.249012501529341E9 & 27.27094413601888 & 1.5572726300028987 & 2.0564420881938667E10 & 1.3507838486974697E9\\
   CANDELS\_GOODSS\_19 & 4.5 & -0.5577911744909068 & 0.10923765003146907 & 1.9773443241802593E-9 & 3.8441244222937E-10 & 0.0013996563208089308 & 2.6132419496119406E-4 & 0.001965034037701186 & 3.6688359159255866E-4 & 0.001224299730424564 & 2.2858407216672831E-4 & 3.3159068447112266E7 & 4073900.7060818397 & 4.655335541183464E7 & 5719513.736800058 & 2.9004719199543502E7 & 3563500.1693485817 & 2.4803586918253133 & 0.13652862903135152 & 1.1193926838989579 & 0.05711257520244856 & 817.4007981199693 & 126.36315212013255 & 5.306601983330469E11 & 6.51928919687117E10 & 4.003982699611688E10 & 2.0019913498058443E9 & 46.84499674256306 & 6.290140237294263 & 2.3690864645935364E10 & 3.330627792062718E9\\
   CANDELS\_GOODSS\_21 & 5.571599960327148 & -1.385195733484872 & 0.2240503999483133 & 2.9611185823600035E-9 & 1.1881331027579991E-9 & 0.001721959860077444 & 6.927449461230134E-4 & 0.002417528993583035 & 9.72572608246038E-4 & 0.0015062233215053138 & 6.059540747125763E-4 & 1.2567260331853388E7 & 1051205.6831745997 & 1.764368434279015E7 & 1475830.114402064 & 1.0992765300821157E7 & 919504.8923064314 & 1.75737511695253 & 0.11196872998014781 & 0.7622250123623201 & 0.05677219649616781 & 605.7674995415812 & 195.00521795928114 & 2.0108653038935873E11 & 1.6826274387839476E10 & 3.470240296958019E10 & 1.7351201484790096E9 & 21.610926573127653 & 1.6945631482413752 & 7.298230709796094E9 & 2.8719145818186316E9\\
   CANDELS\_GOODSS\_32 & 4.4105000495910645 & -0.6785370292415938 & 0.10190585887099374 & 9.7893077452638E-9 & 1.8377886491969703E-9 & 0.005497588714152909 & 9.892909662905158E-4 & 0.0077182868308334415 & 0.001388905542774694 & 0.004808820765966696 & 8.653471894021303E-4 & 3.9707147414796926E7 & 3097332.168897618 & 5.5746467936499976E7 & 4348469.726077503 & 3.4732419061107926E7 & 2709281.4736299426 & 2.4627444624695802 & 0.09101378818920412 & 1.0856434725558535 & 0.04075595672246265 & 231.66836489551997 & 50.424300468198425 & 6.353298506229738E11 & 4.95524890342018E10 & 5.204480724151431E10 & 2.6022403620757155E9 & 70.70472273222722 & 6.68780420350663 & 7.222647869706122E9 & 1.171256705016367E9\\
\end{tabular}

\end{adjustbox}
 \caption{Physical parameters. Only the first rows are shown. For each object in the sample, the table contains the following information for PL+OT\_MBB: id, redshift, $\beta_{FUV}$, sSFR, sM\_dust, M\_dust, A\_FUV,  IRX, age\_main, L\_dust, L\_FUV, SFR, and M\_star.}
 \label{Tab.Physical_parameters_plotmbb}
\end{table}

\sisetup{table-column-width=12ex,    
         round-mode=places,
         round-precision=3,
         tight-spacing,
         table-format = 5.3e-2,
         table-number-alignment = center
         }
  \begin{table}[h!]
  \centering
\begin{adjustbox}{max width=\textwidth}

\begin{tabular}{|l|S|S|S|S|S|S|S|S|S|S|S|S|S|S|S|S|S|S|S|S|S|S|S|S|S|S|S|S|S|l|S|S|S|S|S|S|S|S|S|S|S|S|S|S|S|S|S|S|S|S|S|S|}

\hline
  \multicolumn{1}{|c|}{id} &
  \multicolumn{1}{c|}{redshift} &
  \multicolumn{1}{c|}{$\beta_{FUV}$} &
  \multicolumn{1}{c|}{$\beta_{FUV}$\_err} &
  \multicolumn{1}{c|}{sSFR} &
  \multicolumn{1}{c|}{sSFR\_err} &
  \multicolumn{1}{c|}{sM$_{dust}$} &
  \multicolumn{1}{c|}{sM$_{dust}$\_err} &
  \multicolumn{1}{c|}{A$_{FUV}$} &
  \multicolumn{1}{c|}{A$_{FUV}$\_err} &
  \multicolumn{1}{c|}{IRX} &
  \multicolumn{1}{c|}{IRX\_err} &
  \multicolumn{1}{c|}{age$_{main}$} &
  \multicolumn{1}{c|}{age$_{main}$\_err} &
  \multicolumn{1}{c|}{L$_{dust}$} &
  \multicolumn{1}{c|}{L$_{dust}$\_err} &
  \multicolumn{1}{c|}{M$_{dust}$} &
  \multicolumn{1}{c|}{M$_{dust}$\_err} &
  \multicolumn{1}{c|}{L$_{FUV}$} &
  \multicolumn{1}{c|}{L$_{FUV}$\_err} &
  \multicolumn{1}{c|}{SFR} &
  \multicolumn{1}{c|}{SFR\_err} &
  \multicolumn{1}{c|}{M$_{star}$} &
  \multicolumn{1}{c|}{M$_{star}$\_err} \\
     &   &   &   & [yr$^{-1}$] & [yr$^{-1}$] &   &   &   &   &   &   & [Myrs] & [Myrs] & [L$_\odot$] &  & [M$_\odot$] & [M$_\odot$] & [L$_\odot$] & [L$_\odot$] & [M$_\odot$ yr$^{-1}$] & [M$_\odot$ yr$^{-1}$] & [M$_\odot$ yr$^{-1}$] & [M$_\odot$ yr$^{-1}$] \\
\hline
  CANDELS\_GOODSS\_12 & 4.431000232696533 & -2.144344430007398 & 0.1460967634012939 & 1.1149208811350329E-8 & 3.4331387798514106E-9 & 0.013814630122584031 & 0.004252890556071374 & 1.5888425205667631 & 0.09549578768207853 & 0.6576055662636281 & 0.05054404171237296 & 205.2676153342721 & 68.8026706449075 & 3.289149898984435E11 & 3.444087607084101E10 & 5.488870198868826E7 & 5747427.27130018 & 7.200213239389786E10 & 3.6001066196948934E9 & 44.298370236886015 & 4.647950695210735 & 3.97322993823459E9 & 1.150247287111027E9\\
   CANDELS\_GOODSS\_14 & 5.552700042724609 & -1.9627663677483977 & 0.06621954990145637 & 1.326194276577744E-9 & 1.1576960863836385E-10 & 8.389036769235726E-4 & 1.9276863518511448E-4 & 0.38895337243364636 & 0.07643916190570746 & -0.2178516335164711 & 0.10904574939109347 & 1028.232758327914 & 20.615745087813362 & 1.0336335052519156E11 & 2.2754735569254753E10 & 1.7249077475250736E7 & 3797266.5820972957 & 1.6635877899400333E11 & 8.317938949700168E9 & 27.26847962785562 & 1.5620781687034537 & 2.0561451749152596E10 & 1.3543682219820144E9\\
   CANDELS\_GOODSS\_19 & 4.5 & -0.5577911744909068 & 0.10923765003146907 & 2.0791810501074646E-9 & 4.0873965089761124E-10 & 0.0039209999786145675 & 7.389569536895512E-4 & 2.5099058774851524 & 0.1380529502851573 & 1.1267584398851647 & 0.05772417621057204 & 792.2062394855445 & 126.16545886623689 & 5.49213234346225E11 & 6.8077912733368996E10 & 9.165165004362184E7 & 1.136071136553421E7 & 4.073865421849303E10 & 2.0369327109246514E9 & 48.599942622063324 & 6.609169505728361 & 2.3374560199820686E10 & 3.3182618820359697E9\\
   CANDELS\_GOODSS\_21 & 5.571599960327148 & -1.385195733484872 & 0.2240503999483133 & 2.955331329981361E-9 & 1.1827481700911702E-9 & 0.004595338622398079 & 0.0018443431792643092 & 1.7606691956564664 & 0.11242903814406449 & 0.762352550409809 & 0.057306132909870695 & 606.6511224709545 & 194.58770651261204 & 2.0194602332656482E11 & 1.6969794863164623E10 & 3.370036463097957E7 & 2831886.7843056116 & 3.48423369083178E10 & 1.74211684541589E9 & 21.673210966497958 & 1.6990243294516578 & 7.333597673677641E9 & 2.878110178828852E9\\
   CANDELS\_GOODSS\_32 & 4.4105000495910645 & -0.6785370292415938 & 0.10190585887099374 & 1.018539814396081E-8 & 1.9023428175789387E-9 & 0.015229920938262585 & 0.002726856104238399 & 2.4843473871713986 & 0.09002920142052893 & 1.0898551818463362 & 0.0403862337281479 & 223.19913272814188 & 48.48587187087437 & 6.524372506377365E11 & 5.006973065914676E10 & 1.088774756894815E8 & 8355540.51717848 & 5.293865105835035E10 & 2.646932552917518E9 & 72.81458934042845 & 6.797772856448676 & 7.148919297141285E9 & 1.1564475989098332E9\\
\end{tabular}

\end{adjustbox}
 \caption{Physical parameters. Only the first rows are shown. For each object in the sample, the table contains the following information for PL+OT\_MBB: id, redshift, $\beta_{FUV}$, sSFR, sM\_dust, M\_dust, A\_FUV,  IRX, age\_main, L\_dust, L\_FUV, SFR, and M\_star.}
 \label{Tab.Physical_parameters_dl2014}
\end{table}

\end{document}